\documentclass[fleqn,usenatbib]{mnras}

\usepackage{newtxtext,newtxmath}
\usepackage{longtable}
\usepackage{pdflscape}

\usepackage[T1]{fontenc}
\usepackage{ae,aecompl}

\usepackage{graphicx}	
\usepackage{amsmath}	

\usepackage{amssymb}	
\usepackage{soul}

\title[Photometry of R136 Core]{High contrast and resolution near infrared photometry of the core of R136\thanks{Based on observations made with ESO Telescopes at the La Silla Paranal Observatory under programme ID 0102.D0271}}

\author[Z. Khorrami et al.]{
Zeinab Khorrami,$^{1}$\thanks{E-mail: KhorramiZ@cardiff.ac.uk}
Maud Langlois,$^{2}$
Paul C. Clark,$^{1}$
Farrokh Vakili,$^{3,9}$
\newauthor
Anne S. M. Buckner,$^{4}$
Marta Gonzalez,$^{5}$
Paul Crowther,$^{6}$
Richard W\"unsch,$^{7}$
\newauthor
Jan Palou\v{s},$^{7}$
Stuart Lumsden,$^{8}$
Estelle Moraux$^{5}$\\
$^{1}$School of Physics and Astronomy, Cardiff University, The Parade, Cardiff CF24 3AA, UK\\
$^{2}$ Universite de Lyon, Universite Lyon 1, CNRS, CRAL UMR5574, Saint-Genis Laval, France\\
$^{3}$ Universite Cote d'Azur, OCA, CNRS, Lagrange, France\\
$^{4}$ School of Physics and Astronomy, University of Exeter, Stocker Road, Exeter, EX4 4QL, UK\\
$^{5}$ Universite Grenoble Alpes, CNRS, IPAG, F-38000 Grenoble, France\\
$^{6}$ Department of Physics and Astronomy, Hounsfield Road, University of Sheffield, Sheffield, S3 7RH, UK\\
$^{7}$ Astronomical Institute of the Czech Academy of Sciences, Bo\v{c}n\'\i II 1401/1a, 141 00 Praha 4, Czech Republic\\
$^{8}$ School of Physics and Astronomy, University of Leeds, Leeds LS2 9JT, UK\\
$^{9}$ Department of Physics, Shahid Beheshti University, G.C., Tehran, Iran
}

\date{Accepted XXX. Received YYY; in original form ZZZ}

\pubyear{2020}

\begin{document}
\label{firstpage}
\pagerange{\pageref{firstpage}--\pageref{lastpage}}
\maketitle

\begin{abstract}
We present the sharpest and deepest near infrared photometric analysis of the core of R136, a newly formed massive star cluster at the centre of the 30 Doradus star forming region in the Large Magellanic Cloud. We used the extreme adaptive optics of the SPHERE focal instrument implemented on the ESO Very Large Telescope and operated in its IRDIS imaging mode, for the second time with longer exposure time in the H- and K filters. Our aim was to (i) increase the number of resolved sources in the core of R136, and (ii) to compare with the first epoch to classify the properties of the detected common sources between the two epochs.
Within the field of view (FOV) of $10.8"\times12.1"$  ($2.7\rm{pc}\times3.0\rm{pc}$), 
we detected 1499 sources in both H and K filters, for which 76\% of these sources have visual companions closer than 0.2".
The larger number of detected sources, enabled us to better sample the mass function (MF). The MF slopes are estimated at ages of 1, 1.5 and 2 Myr, at different radii, and for different mass ranges.
The MF slopes for the mass range of 10-300 M$_\odot$ are about 0.3 dex steeper than the mass range of 3-300 M$_\odot$, for the whole FOV and different radii.
Comparing the JHK colours of 790 sources common in between the two epochs, 67\% of detected sources in the outer region ($r >3"$) are not consistent with evolutionary models at $1-2$Myr and with extinctions similar to the average cluster value, suggesting an origin from ongoing star formation within 30 Doradus, unrelated to R136.
\end{abstract}




\section{Introduction}

R136 is a very massive young star cluster that lies at the centre of the Tarantula nebula in the Large Magellanic Cloud (LMC). Hosting the most massive stars known in the Local Universe \citep{crowther2010, crowther2016}, R136 provides a unique opportunity to observationally study the formation of massive stars and clusters in the earliest stages of their evolution. Our understanding of the true nature of R136 has significantly improved with increasing telescope resolution.
The combination of photometry \citep{hunter95,Brandl96,andersen2009, demarchi2011,Cignoni2015}, ultraviolet spectrometry (STIS/MAMA, \cite{crowther2016}), visible (HST/FOS, \cite{massey98}) and near infrared (NIR) observations (VLT/SINFONI, \cite{schnurr2009}) has resulted in more constraints on the R136 stellar population and its most luminous stars.

In 2015, R136 was observed for the first time in the NIR by the second generation Spectro-Polarimetric High-contrast Exoplanet Research {\footnote{\url{https://www.eso.org/sci/facilities/paranal/instruments/sphere.html}}} (SPHERE, \citealt{sphere}) instrument of the Very Large Telescope (VLT). 
Thanks to SPHERE's high contrast and extreme adaptive optics (XAO), the sharpest images from the central region of R136 were recorded in J and K. For the first time, more than one thousand sources were detected in these bandpasses within the small field of view (FOV) of IRDIS ($10.9"\times12.3"$) covering almost $2.7\times3.1$ pc of R136 core \citep{khorrami17}.
The SPHERE data on R136 core were used to partially resolve and study the initial mass function (IMF) covering a mass range of 3 to 300 M$_\odot$, at ages of 1 and 1.5 Myr. 
The density in the core of R136 ($r < 1.4$~pc) was estimated and extrapolated in 3D for larger radii (up to 6pc).
Even at high angular resolution in the NIR the stars in the core are still unresolved due to crowding and central concentration of bright sources.
Using evolutionary models of stars more massive than 40 solar masses, \cite{Bestenlehner2020} finds that the initial mass function (IMF) of massive stars in R136 is top-heavy with a power law exponent $\gamma=2.0\pm0.3$. They also estimate the age of R136 between 1 and 2 Myr with a median age of around 1.6 Myr.

This paper presents the second epoch observation of R136 in the NIR by VLT/SPHERE. These observations, obtained with longer exposures in different NIR filters, aimed first to increase the number of resolved sources in the core of R136, and second to enable comparison with the first epoch to classify the property of the detected common sources between these two epochs.
We divided this paper as follows. In Section \ref{sec:observations} we explain in detail the data and the observing conditions. 
We then present the photometric analysis in Section \ref{sec:photometry} and the method we use to correct for completeness in Section \ref{sec:com}.
In Section \ref{sec:extinction} we describe the details of extinction estimation. 
On the basis of the results we investigate the stellar mass functions and the density in Section \ref{sec:mass}.
In Section \ref{sec:jhk}, we compare the photometric analysis between the two epochs and between the two sets of filters. 
We conclude by summarising our results in Sec. \ref{sec:summary}.

\section{Observations}\label{sec:observations}

We obtained the data via ESO Time Observation run (0102.D-0271) to image R136 using the ``classical imaging" mode of IRDIS \cite{maud14}
in H- and K-bandpass filters.
For our purpose we used the same spectral band split into the two IRDIS channels to correct for residual detector artefacts such as hot pixels and uncorrelated detector noise among other instrumental effects. 
We maximized the image sharpness of the total exposure by discarding the single frames with poorer Strehl ratio and by correcting "a posteriori" the residual tip-tilt image motion on each short exposure before combining them. 
Observations were performed in October/November (K/H) 2018, achieving high dynamical range and high angular resolution imaging in both the H and K bandpass, over a FOV of $10.8"\times12.1"$ ($2.7\rm{pc}\times3.0\rm{pc}$), centered on the core of the cluster. The natural seeing was $0.69 \pm 0.10$ arcseconds in K and $0.58 \pm 0.05$ in H during the observations. 
The night was rated as ``Clear" meaning that less than 10\% of the sky (above 30 degrees elevation) was covered by clouds, and the transparency variations were less than 10\% during the exposures.

\begin{figure*}
    \centering
    \includegraphics[trim=0 0 0 0,clip,width=16cm]{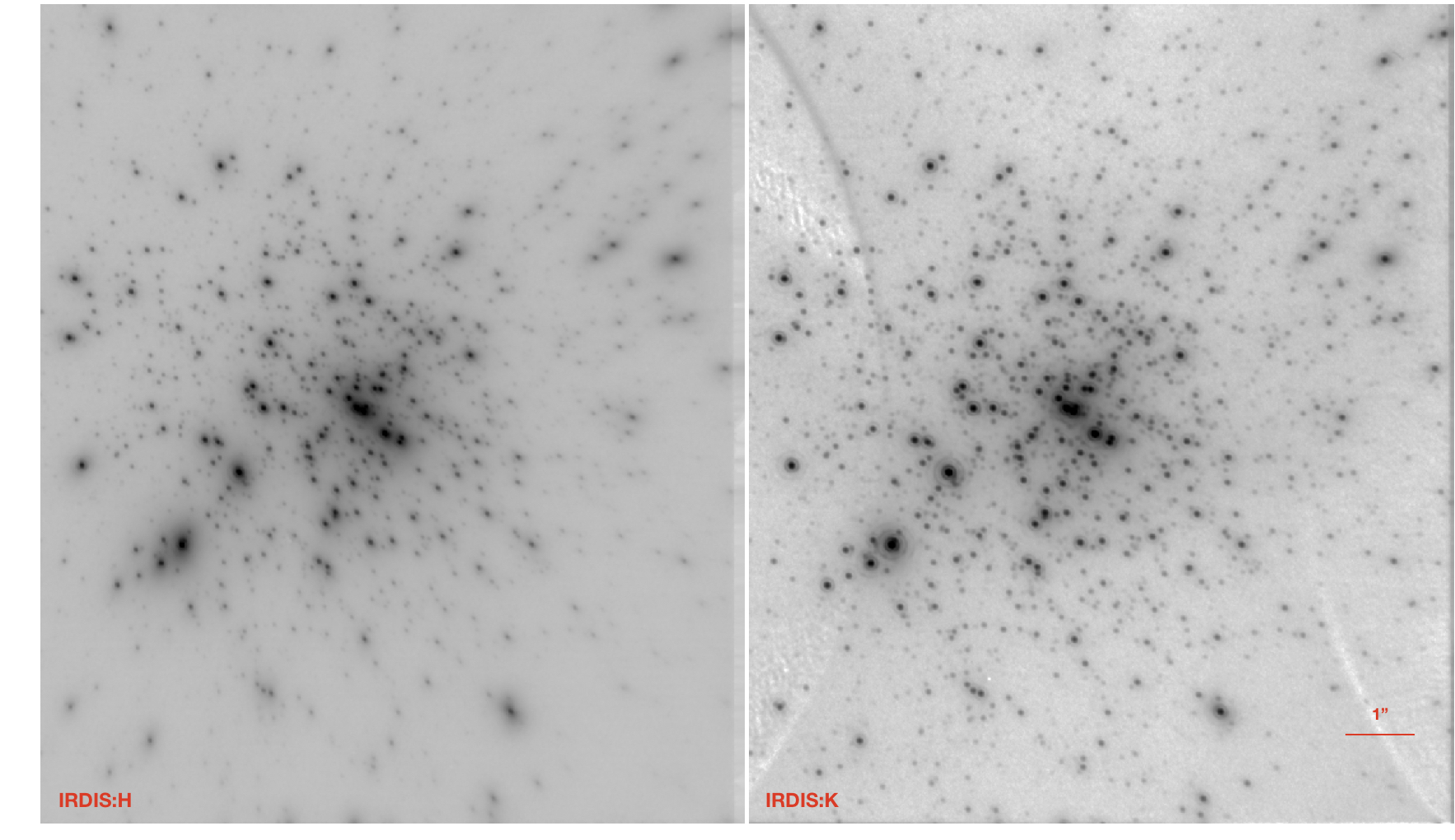}
    \caption{R136 core images taken by VLT/SPHERE/IRDIS in the H (left) and K (right). FOV is $10.8"\times12.1"$.
    }
    \label{fig:irdisimg}
    
\end{figure*}

\begin{figure*}
    \centering
    \includegraphics[trim=0 0 0 0,clip,width=16cm]{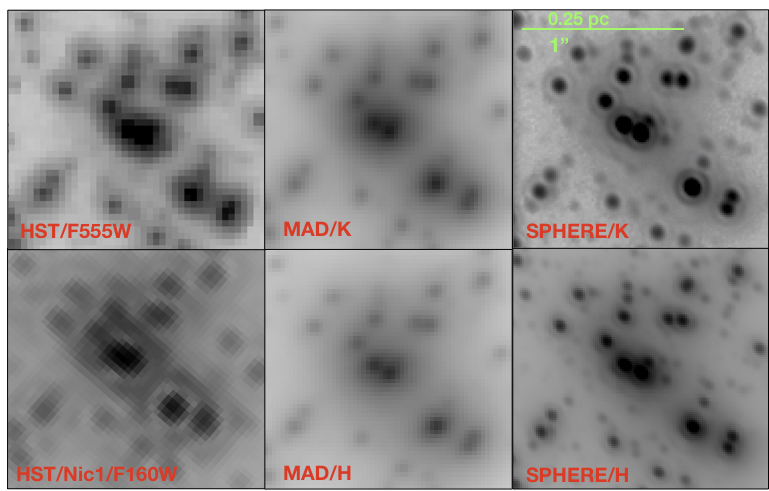}
    \caption{Comparison of R136 central images at different wavelengths with the highest available angular resolution telescopes. Top, left to right: HST/WFPC2 in V , VLT/MAD in K, and VLT/SPHERE in K. Bottom, left to right: HST/NICMOS/NIC1, VLT/MAD, and VLT/SPHERE in the H. The identification of bright sources is shown in Figure \ref{extra:hunter}.
    }
    \label{fig:coreimg}
    
\end{figure*}

Our data consists of 1088/544 frames of 2.0/4.0s exposures taken with the IRDIS broad-band H and K filters (BB-H, BB-K). 
Table \ref{table:expo} shows the exposure time log of the two epoch observations and the filters information.
Figure \ref{fig:irdisimg} shows the final reduced images in the H (left) and K (right). These images are the deepest and sharpest images taken from the core of R136 so far.
The Wolf-Rayet star R136a1 was used for guiding the AO loop of SPHERE confirming the high level of performances even for faint guide stars which surpass both NACO and MAD performances.
Figure \ref{fig:coreimg} compares the highest resolution available images from the very central part of R136 taken by HST (WFPC2 by \citealt{hunter95}) in left (top in V, bottom in H), VLT/MAD \citep{Campbell2010} in the middle (top in K, bottom in H), and IRDIS in right (top in K, bottom in H). These images shows the effect of better contrast, pixel-sampling, sensitivity, and AO correction on resolving the stellar sources in the most crowded part of R136.

The range of airmass during these observations was 1.52 to 1.45 in K, and 1.70 to 1.55 in the H. A log of observations is provided in Table \ref{table:exp2}. 
For comparison we include the observing log for our first epoch data-sets taken in 2015.

\begin{table}
\caption{Exposure time log of VLT/SPHERE observations on R136.}
\centering
\begin{tabular}{ c c c c c  c}  
Obs. date& Filter&Single/Total &$\lambda_{cen}$[nm] &$\Delta\lambda$ [nm] \\
    & & Exposure[s] && \\
\hline
2015-09-22 &BB-J&4.0/1200 &1245&235\\
2015-09-22 &BB-K&4.0/1200 &2182&294\\
\hline
2018-10-10 &BB-K&4.0/4352 &2182&294\\
2018-11-14 &BB-H&2.0/4352 &1626&286\\
\end{tabular} 
\label{table:expo}  
\end{table}

\begin{table}
\caption{Observing condition for R136 data in 2015 and 2018}
\centering
\begin{tabular}{ l c c c c  c}  
Data &SR &Seeing["]& Airmass & N$_{phot}$\\
\hline
2015-J&$0.40\pm0.05$&$0.63\pm0.1$&1.54-1.59&1110\\
2015-K&$0.75\pm0.03$&$0.63\pm0.1$&1.61-1.67&1059\\
\hline
2018-H&$0.71\pm0.05$&$0.58\pm0.05$&1.70-1.55&1658\\
2018-K&$0.83\pm0.03$&$0.69\pm0.10$&1.52-1.45&2528\\
\end{tabular} 
\label{table:exp2}  
\end{table}

We used the SPHERE pipeline package to correct for dark (current), flat (fielding), (spatial) distortion, bad pixels and thermal background due to instrument and sky (in K). Furthermore and in order to reach the highest sensitivity and the largest number of detectable sources, additional corrections were carried out on the images. Based on a Gaussian fit using selected stars, we estimated and corrected the subpixel image drifts before combining the individual images. This allowed us to correct for residual tip-tilt errors with an accuracy of a few mas. The final step is performed to combine the parallel images on the left and right side of the detector, after the anamorphism correction.  Some uncorrected background leaks persist in our final K images due to thermal background fluctuations, which are stronger in K than in H.

\section{Photometry}\label{sec:photometry}
To analyse the final images we applied the same method/tools as on first epoch data \citep{khorrami17}.
For the photometry we used the {\it{Starfinder}} package implemented in IDL \citep{starfinder}. {\it{Starfinder}} is designed for the analysis of AO images of crowded fields, like the Galactic Center, for instance,  as in \cite{Pugliese2002}. 
This method determines the empirical local Point Spread Function (PSF) from several isolated sources in the image and uses this PSF to extract other stellar sources across the FOV.

For the present study, four and seven well isolated stars (within $0.47"\times0.47"$) were used to extract the PSF in the K and H, respectively. 
The FOV in K is more crowded than in H (see Figure \ref{fig:irdisimg}), so finding the bright isolated stars (within $0.47"\times0.47"$) in K is harder than in H. That is why we limit our selection of stars to be used for PSF extraction, to four in K. Figure \ref{extra:sep} shows the separation with the closest neighbour for each star in H and K.
The extracted PSFs, were used as an input for stellar sources detection by {\it{Starfinder}}. 
The FWHM of the extracted PSFs are 58.8 and 63.7 mas in H and K, with 1658 and 2528 sources detected, respectively.
We stopped the source extraction after obtaining a minimum correlation coefficient of 65\% and 80\%, in K and H, between the extracted star with the locally determined PSF according to {\it{Starfinder}} procedures. 
In H, we put a higher threshold on minimum correlation coefficient than in K, because the AO correcting efficiency in H degrades faster as a function of distance from R136a1 -which is the
reference star for the AO loop- than in K. The isoplanatic angle in H is smaller than in K, so at larger radii from R136a1, the PSF is not centrosymmetric.
Indeed, stars with higher correlation coefficients, i.e. more similarity to the PSF, represent higher reliability on their photometry estimation. 

\begin{figure}
    \centering
    \includegraphics[trim=0 0 0 0,clip,width=8.2cm]{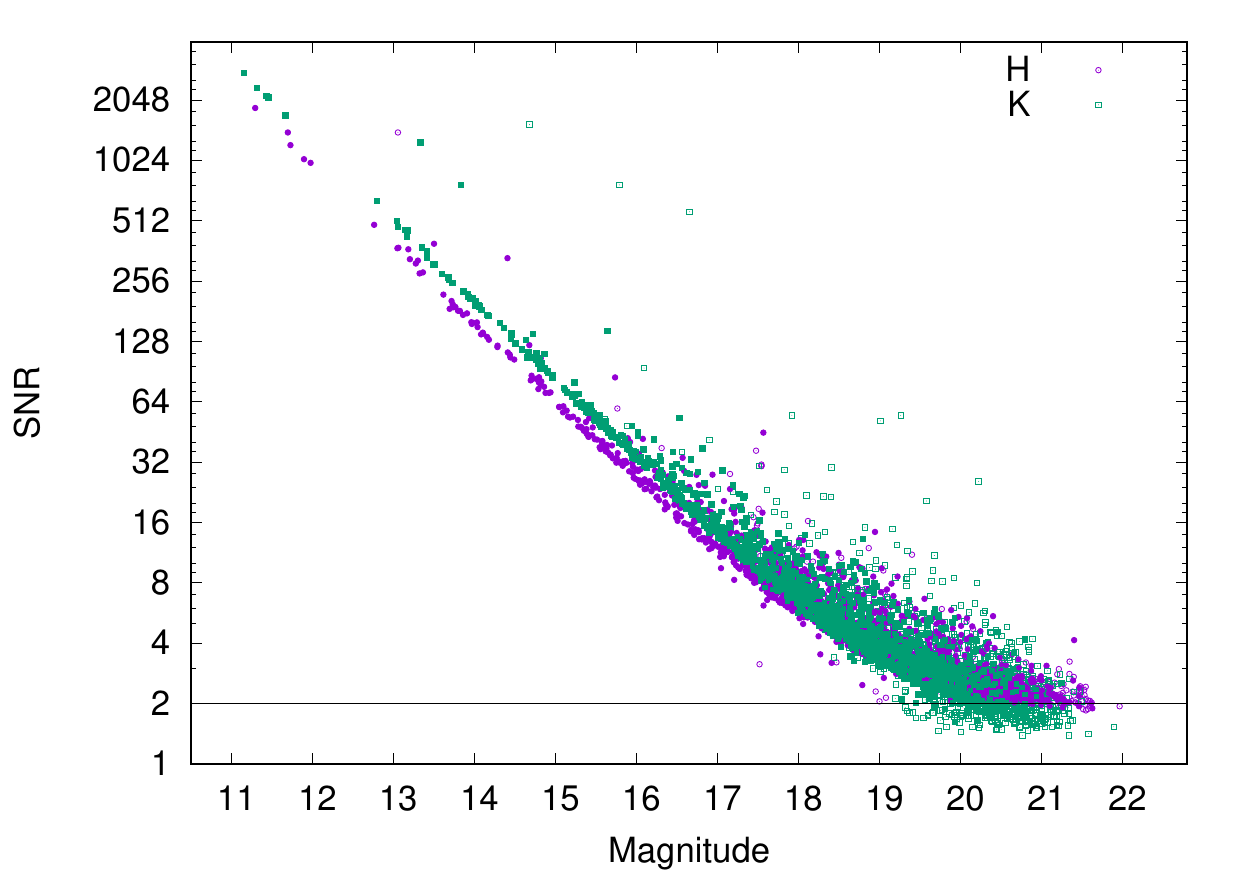}
    \caption{Signal to noise ratio of detected sources in H (purple circles) and K (green squares) versus their magnitudes. Filled dots are common sources between H and K data. The solid horizontal line shows the SNR=2.0.}
    \label{fig:snr}
\end{figure}

\begin{figure}
    \centering
    \includegraphics[trim=0 0 0 0,clip,width=8.2cm]{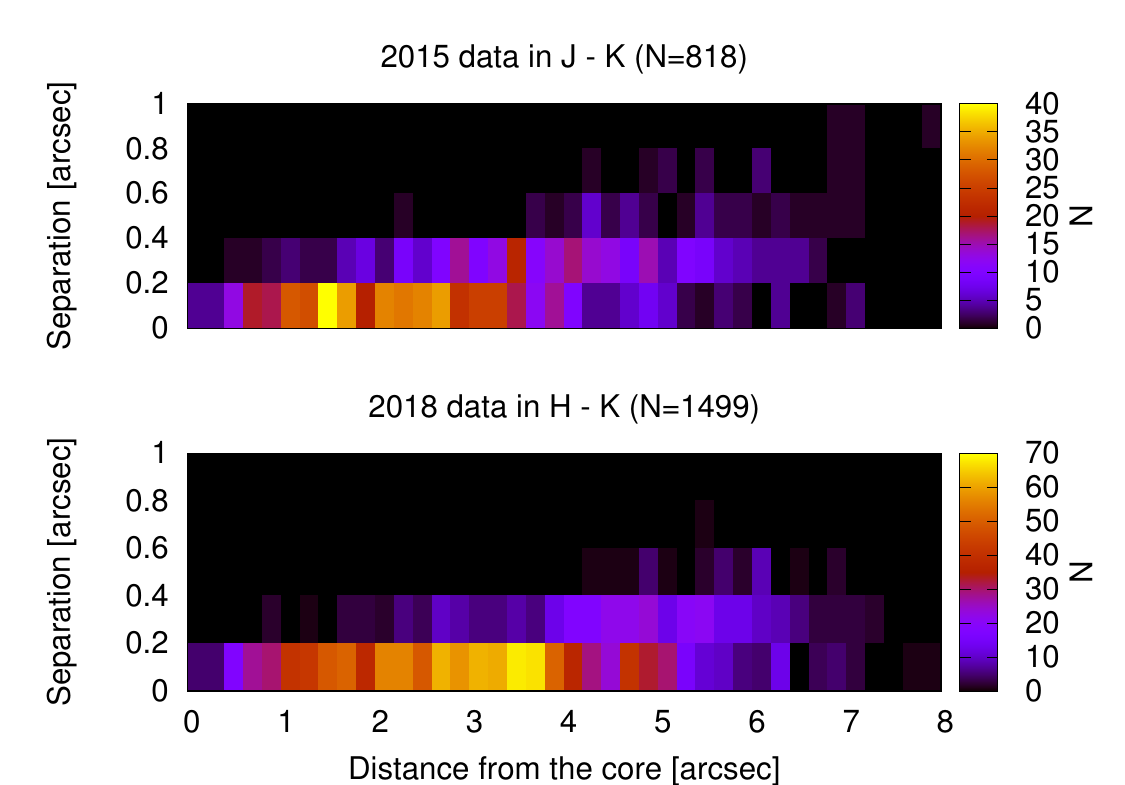}
    \caption{Histogram of the separation of the close detected sources versus their distance from the core of R136. Top: 818 common sources between J and K data in 2015. Bottom: 1499 common sources between H and K data in 2018.}
    \label{fig:sep}
\end{figure}

In addition to the correlation coefficient criterion, we applied the limit of standard deviation from the sky brightness ($\sigma_{sky}$) for stopping the extraction of sources by {\it{Starfinder}}, i.e. the local PSF maximum value must exceed $2\sigma_{sky}$ over the background. 
Figure \ref{fig:snr} shows the signal to noise ratio (SNR) of 1658 and 2528 detected sources in H (purple circles) and K (green squares), respectively. Common sources (1499) between H and K are shown as filled circles/squares. 
The solid horizontal line shows the SNR=2.0 where some of the detected sources have SNR less than this value. These stars are located further from the central region where the local $\sigma_{sky}$ is smaller than the $\sigma_{sky}$ of the whole FOV.
76\% of these sources have visual companions closer than 0.2" which is much higher than the value found in the first epoch in 2015 data. Figure \ref{fig:sep} shows the histogram of closest neighbour (visual companion) separation from each source versus its distance from the centre of R136, in 2015 (top) and 2018 (bottom) data. 
Comparing the histogram of closest companions between two epochs, 85\% of the new detected sources in 2018 have visual companions with separation less than 0.2". 

The Strehl Ratio (SR) in H and K is determined as $0.71 \pm 0.05$ and $0.83 \pm 0.03$, respectively, on average within 5 arcseconds from the core. These estimates are assessed from the SPARTA files recorded during SPHERE runs simultaneously with the AO corrected images of R136. 
They are based on the slope measurements of the Shack-Hartmann wavefront sensor in Sphere AO for eXoplanet Observation (SAXO), by extrapolating the phase variance deduced from the reconstruction of SAXO open-loop data using deformable mirror, tip tilt voltages and wavefront sensor closed-loop measures \citep{fusco}.
The method has been proved quite robust for FOV smaller than the anisoplanetism in the past for differential imaging by NACO \citep{maire}.
We are therefore quite confident on our photometry corrected for the SR effect in H and K.
To convert stellar fluxes to magnitudes, we set the zeropoint magnitude to the instrumental zeropoint (one ADU/s in H and K, are 25.2 and 24.8 magnitudes, respectively). 
To double check our calibration, we compared H and K magnitudes of 26 sources (located $r \geq 3"$ from R136a1) common between our catalogue and the VLT/MAD \citep{Campbell2010} in H and K {\footnote{See the comparison between two catalogues in Figure \ref{extra:mad}.}}.
We cannot compare our zeropoints with the MAD/H and K magnitudes in the central 2.8" region because of the completeness issue and the very low AO correction (SR $\sim$ 15-30\%) of their data. Figure \ref{fig:coreimg} shows the completeness and contrast problem of the previous data, leading to an inability to distinguish certain sources. 
By way of example R136a6 (H19) and H26{\footnote{Identification from Hunter et al. (1995) - see Figure \ref{extra:hunter} to check the position of these stars in the core.}} - are very close in terms of brightness (flux ratio of about 90\%) and location (Separation of about 70mas). 
These bright O-type stars are most easily detectable in our data compared with others (Figure \ref{fig:coreimg} and Figure \ref{extra:hunter}). Cases like these close bright stars, bring confusion even in the HST spectroscopic analysis \citep{Bestenlehner2020}.

The K magnitude of R136a1 and R136a2 (11.15~mag and 11.43~mag) increased about 0.08 and 0.11 mag, compare to their K-mag in the first epoch (11.07~mag and 11.32~mag). The brightness of R136a3 and R136b in K (11.45~mag and 11.67~mag) are consistent with the first epoch K data (11.44~mag and 11.66~mag).
R136c in the K (11.31~mag) is 0.17~mag brighter than in 2015 (11.48~mag), making this source the second brightest in our K catalogue. In previous studies R136c shows indications of binarity based on its strong X-ray emission \citep{zwart2002,townsley2006,guerrero2008} and possible low-amplitude RV variations \citep{schnurr2009}.

Comparing the brightness of common detected sources within two epochs, the K magnitude of 56\% of the sources have changed more than twice their photometric errors 
{\footnote{see Figure \ref{extra:KKcorr} to compare the K~mag of these sources within two epochs.}}. 
To make sure that this is not due to {\it{Starfinder}} photometry analysis (which is based on the reference input PSF), we have used {\it{DAOPHOT}} algorithm for this analysis as well. The variation of the K~magnitude of these sources still remains 
{\footnote{see Figure \ref{extra:KKfind} to compare the photometric analysis using {\it{Starfinder}} and {\it{DAOPHOT}}}}.
The K~mag of about 67\% of the sources located outside of the core ($r > 3"$), has changed more than twice their photometric errors. 
Since the observations were made in different nights and the AO correction in the outer part of the FOV is not as good as its center, this inconsistency is probably an observational effect, otherwise it is originated from physical reasons like variables or multiple systems with periods less than 3 years.

\section{Completeness}\label{sec:com}
\begin{figure}
    \centering
    \includegraphics[trim=0 0 0 0,clip,width=8.2cm]{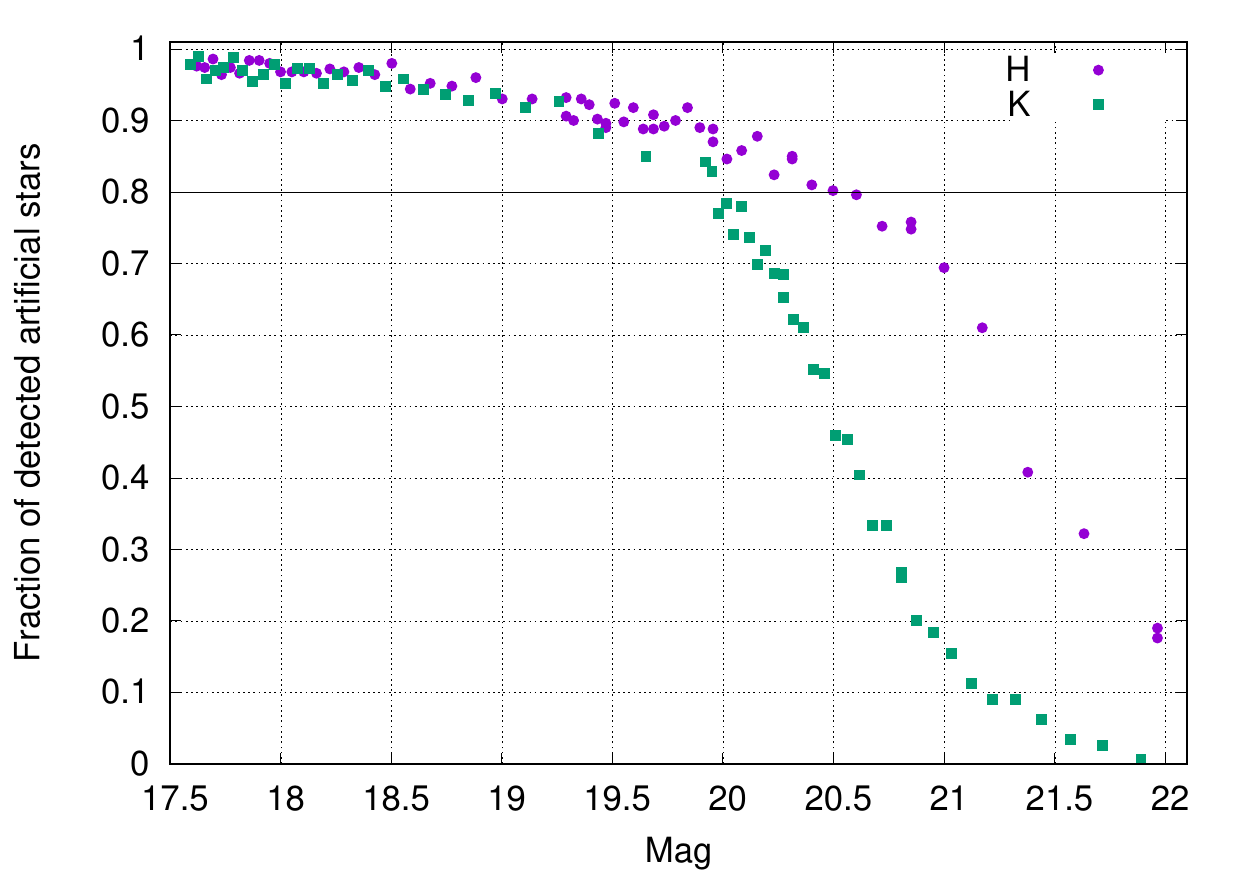}
    \caption{Average value of completeness values for the whole FOV as a function of magnitude using 62,000 artificial stars in H (purple dots) and K (green squares).}
    \label{fig:com}
\end{figure}

We produced $6.2\times10^4$ artificial stars (500 per magnitude) in H and K, in order to determine the completeness as a function of magnitude. 
These artificial stars are created using the same PSF as we used for source extraction process. 
They are added to the original image one by one (500 times per magnitude) and the same photometric tools and criteria to determine how often they can be recovered from the source extraction process are used. These tests were performed one star at a time to avoid
the artificial stars from affecting one another.
Figure \ref{fig:com} shows the completeness values as a function of magnitude in H (purple dots) and K (green squares). The completeness is 80\% at $K=20$ mag and $H=20.5$ mag.

\begin{figure}
    \centering
    \includegraphics[trim=40 0 50 0,clip,width=8.2cm]{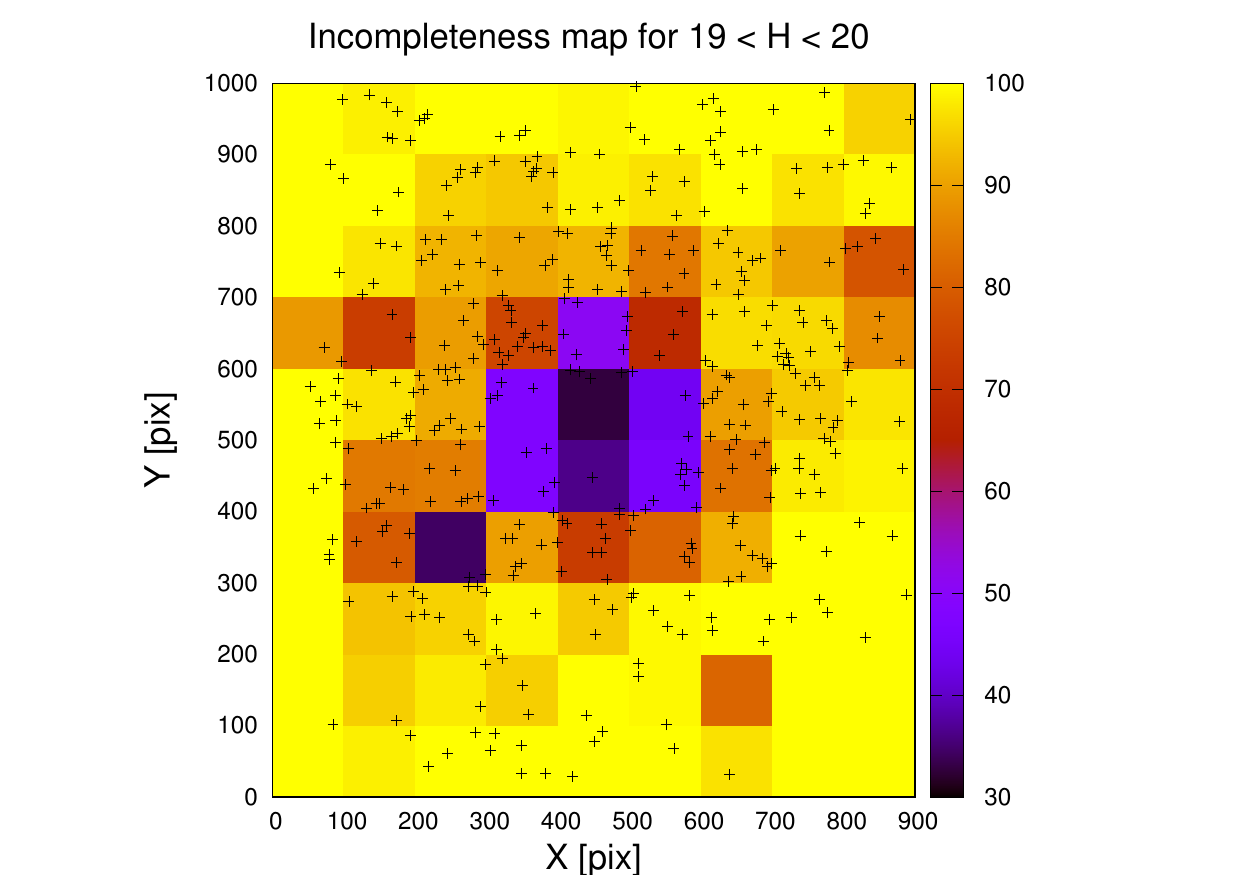}\\
    \includegraphics[trim=40 0 50 0,clip,width=8.2cm]{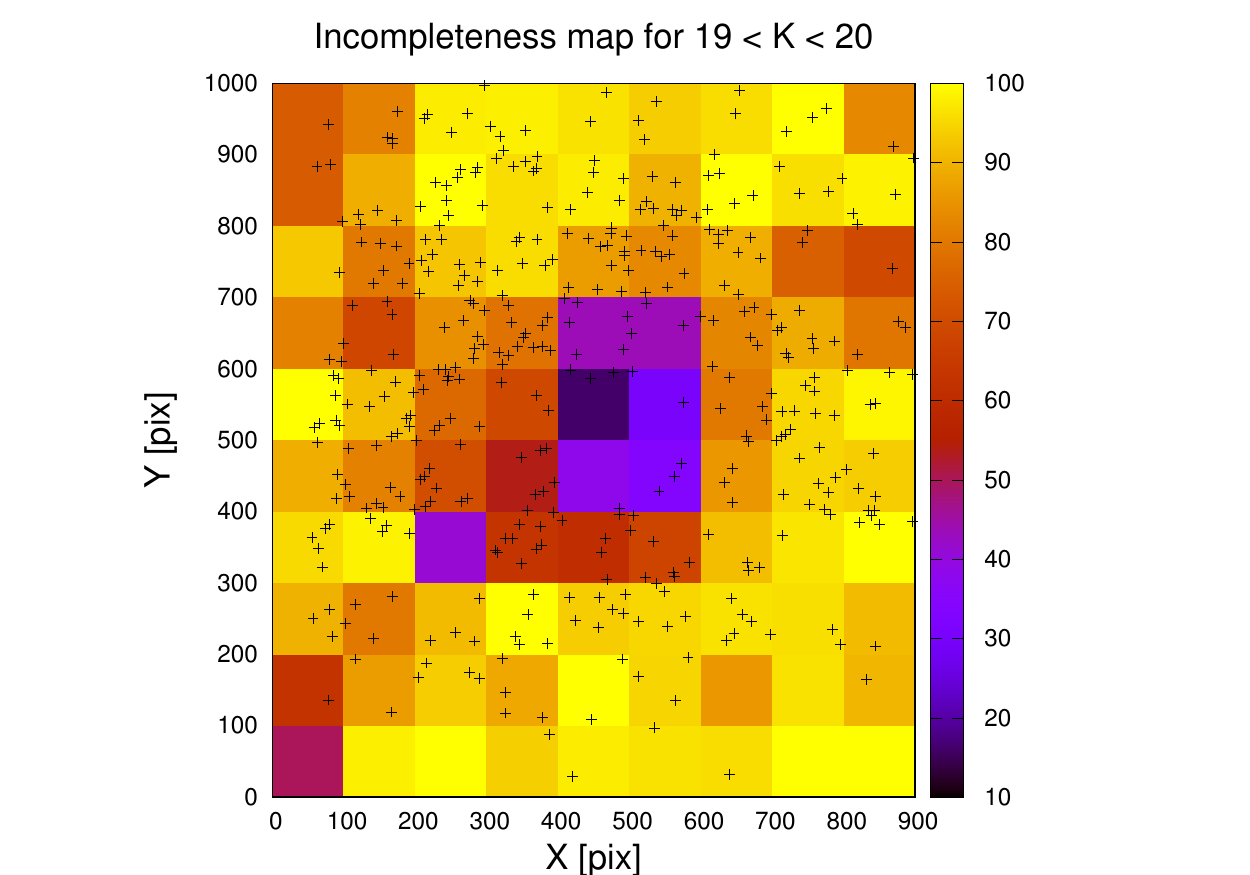}
    \caption{Completeness map for magnitude range of 19 - 20 in H (top) and in K (bottom). Black pluses shows the position of photometric detected sources within the given magnitude range. colour shows completeness in percents.
    }
    \label{fig:commap}
\end{figure}

The completeness value in each magnitude is the average value for detection of 500 artificial stars which are distributed randomly in the FOV. But the completeness varies across the FOV depending on how crowded and bright that region is. 
Using these $6.2\times10^4$ artificial stars, we created completeness maps covering magnitude range of 17 to 22 (5 maps), both in H and in K. Figure \ref{fig:commap} shows two examples of completeness map in the magnitude of 19 to 20 both in H (top) and K (bottom). The average value of completeness is about 90\% both in H and K (Figure \ref{fig:com}), but one can notice that the completeness in the very central part of the cluster is much lower than the outskirts.
In the K we can also see the effect on the completeness of the thermal background noise from the instrument lying in the arcs located in the far east and west parts of the FOV as seen on Figure \ref{fig:commap}.
The black pluses in Figure \ref{fig:commap} shows the position of the observed sources (398 in H and 410 in K) within the magnitude range of 19 to 20. This plot clearly shows the effect of completeness on the detected sources which are used to create MF. To overcome this limitation we corrected the MF for all stars fainter than 17 magnitude. 
Each star contributes a mass distribution to the MF (see Section \ref{sec:mass}). When we added stars with a given mass to the histogram of mass, we corrected the contribution of each star to the mass function (fainter than 17 mag) for completeness using the completeness maps in every H and K magnitudes. We know that the completeness in a given magnitudes varies across the FOV in both filters, so depending on the position and magnitude of an observed star, we estimated its completeness using these maps (see Figure \ref{fig:commap}). 
Since the average value of completeness at 17~mag is above 95\% (above 80\% at the core), we stopped the completeness correction for sources brighter than K=17~mag.

\section{Extinction} \label{sec:extinction}

In order to fit the evolutionary models to our data and estimate the stellar masses, we need to measure the extinction first. We used two methods to estimate the extinction for the spectroscopically-observed massive stars in the core of R136
{\footnote{List of spectroscopically-known stars are provided in Table \ref{table:infostar}. List of stars brighter than 16~mag in K (138 sources) are provided in Table \ref{table:bright16info} with their SPHERE (J, H, K) and HST (U, V, B) magnitudes. See Figure \ref{extra:hunter} to check the position of these sources in our data.}}. 
In both methods, we adopt the LMC distance modulus (DM) of 18.49 magnitude estimated by \cite{pietrzyski2013} which is consistent with the value suggested by \cite{gibson2000} for LMC.

\begin{figure}
    \centering
    \includegraphics[trim=0 0 0 0,clip,width=8.2cm]{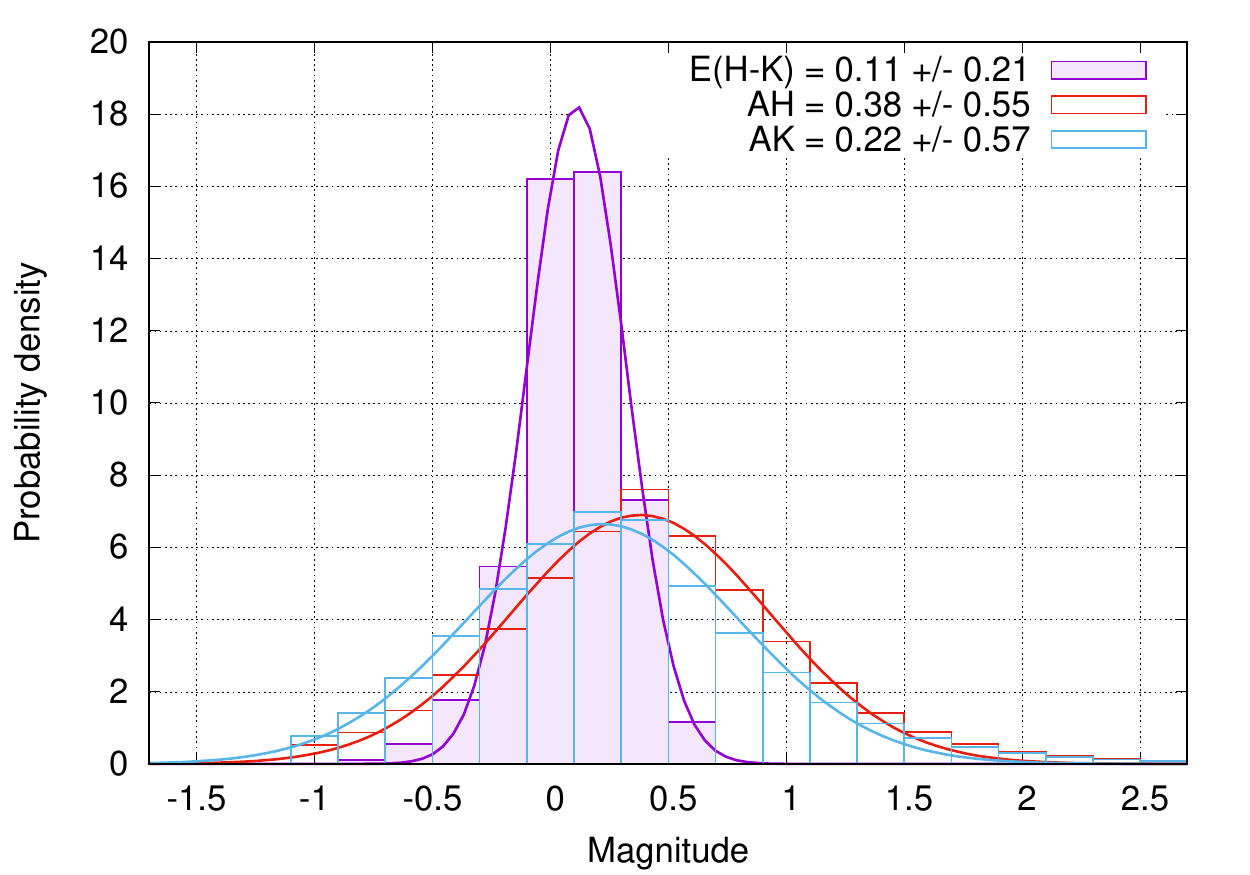}
    \caption{Histogram of extinction in H and K filters from observations using 49 spectroscopically known stars (Bestenlehner et al. 2020) in the FOV.}
    \label{fig:ext}
\end{figure}

\textit{Method I:} We used the effective temperature (T$_{\rm{eff}}$) and luminosity (logL) of 49 stars by \cite{Bestenlehner2020} in the optical. 
The values of these temperatures and luminosities are provided in Table \ref{table:infostar}. 
We also chose a grid of isochrones at different ages (from 0.1 up to 10 Myr) with the LMC metallicity (Z=0.006), from the latest sets of PARSEC evolutionary model \footnote{\url{http://stev.oapd.inaf.it/cgi-bin/cmd}, YBC version of bolometric corrections as in \cite{chen2019}} \cite{Bressan12} 
which is a complete theoretical library that includes the latest set of stellar phases from pre-main sequence to main sequence and covering stellar masses from 0.1 to 350 M$_{\odot}$. 
By fitting the PARSEC isochrones to the observed stellar parameters of each star (T$_{\rm{eff}}$, logL), we estimated the age and intrinsic colour for each star with an error. The extinction in H and K (A$_H$ and A$_K$) are estimated by comparing these intrinsic H and K magnitudes with the observed values from our catalogue.
Figure \ref{fig:ext} shows the histogram of the extinction values of those 49 stars. 
The mean extinction in H and K, and the mean colour excess are A$_H=0.38\pm0.55$, A$_K=0.22\pm0.57$, and E(H-K)$=0.11\pm0.21$.
These values derived by fitting a Gaussian function on the extinction distributions, while these distributions are not exactly Gaussian and symmetric. That is why the mean colour excess is slightly (0.05~mag) higher than A$_H - \rm{A}_K$.
The large errors on these values results from the errors in the reported stellar parameters (T$_{\rm{eff}}$ and log L) from the spectroscopic analysis in the optical (see Table \ref{table:infostar}).

\begin{figure}
    \centering
    \includegraphics[trim=0 0 0 0,clip,width=8.2cm]{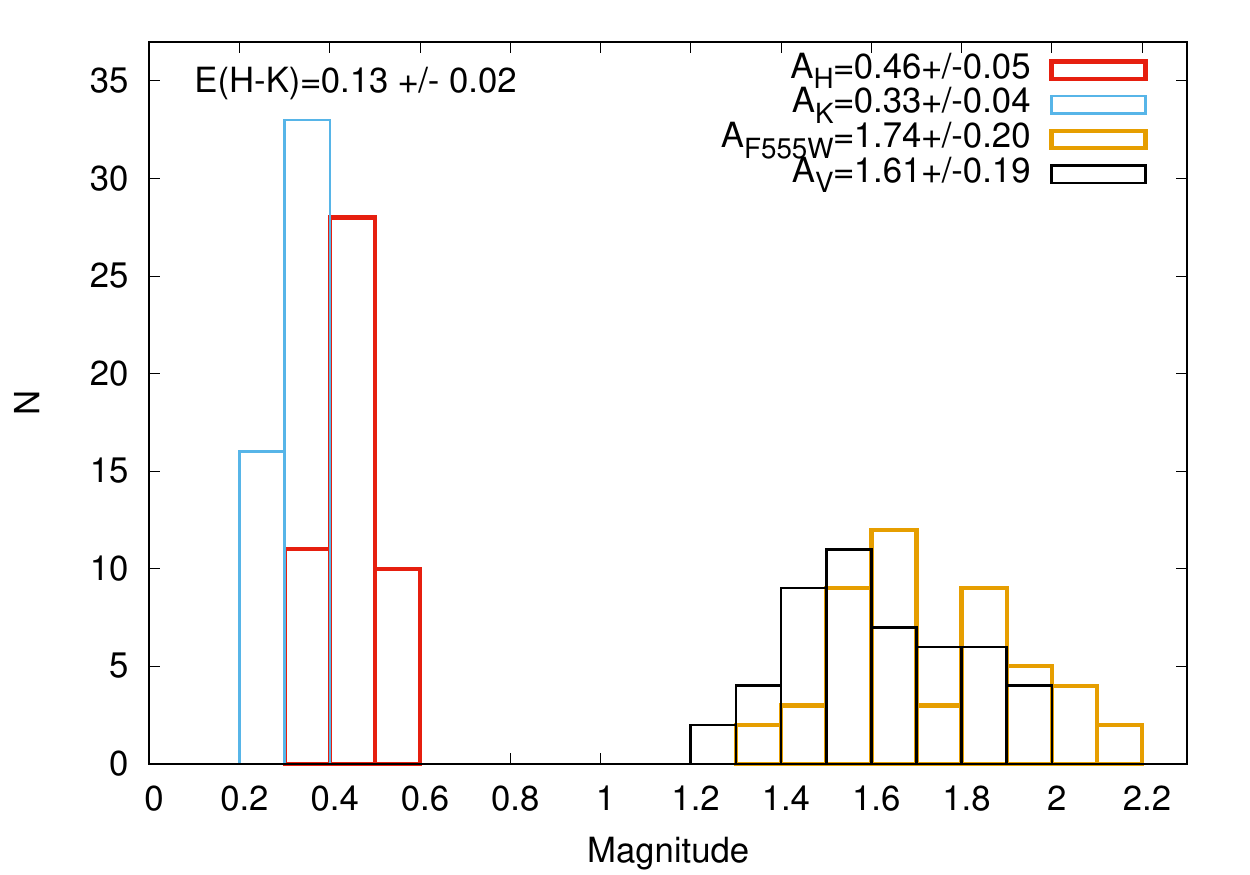}
    \caption{Histogram of extinction in H and K filters modeled based on observed stellar parameters of T$_{\rm{eff}}$, logL and logg from Bestenlehner et al. (2020) and $A_{F555W}$ from Crowther et al. (2016)}
    \label{fig:ext2}
\end{figure}

\textit{Method II:} Using the stellar parameters from \cite{Bestenlehner2020}, we adopt the closest stellar atmosphere model of that star. 
We used the grids of TLUSTY \footnote{\url{http://nova.astro.umd.edu/Tlusty2002/tlusty-frames-cloudy.html}} atmosphere models \citep{tlusty} for O-type \citep{tlustyO} stars with the LMC metallicity ($Z=1/2 Z_\odot$).
The apparent magnitudes of these stars in the SPHERE/IRDIS H and K filters are calculated by MYOSOTIS \footnote{Star cluster simulator tool \url{https://github.com/zkhorrami/MYOSOTIS}} \citep{myosotis}. 
We adopt A$_{F555W}$ for these sources from \cite{crowther2016}, as an input for MYOSOTIS to estimate the H and K magnitude of these stars using the synthetic extinction curves \footnote{\url{www.astro.princeton.edu/~draine/dust/dustmix.html}} from \cite{drainea, draineb, drainec, LD01, WD01} for R$_V=4.0$. 
The simulated H and K magnitudes of 55\% (85\%) of these sources (except for three WRs) are less than 0.1 (0.2) magnitudes different from the observed values. One can compare the result of these simulation with observation in Figure \ref{fig:cmdR3}-top, where black circles are 49 spectroscopically-known stars from \cite{Bestenlehner2020} and black squares are their simulated magnitudes. 
The simulated extinction and colour excess of these 49 stars in H and K are 
A$_H=0.46\pm0.05$, A$_K=0.33\pm0.05$, and E(H-K)$=0.13\pm0.02$, shown in Figure \ref{fig:ext2}.

The colour excess, E(H-K), estimated by two methods are consistent,
and the extinction values, A$_H$ and A$_K$, in \textit{Method II} are 0.1\,mag higher than \textit{Method I}. 
We adopt the extinctions of A$_H=0.45$ and A$_K=0.35$ magnitude, which are consistent with the two analyses (within the error-bars) and \cite{demarchi14}. 
The colour excess of $E(H-K)=0.1$ is also consistent with \cite{tatton} and with the one used in previous studies by \cite{Campbell2010}.

\section{Mass Function and core density}\label{sec:mass}
\begin{figure*}
    \centering
    \includegraphics{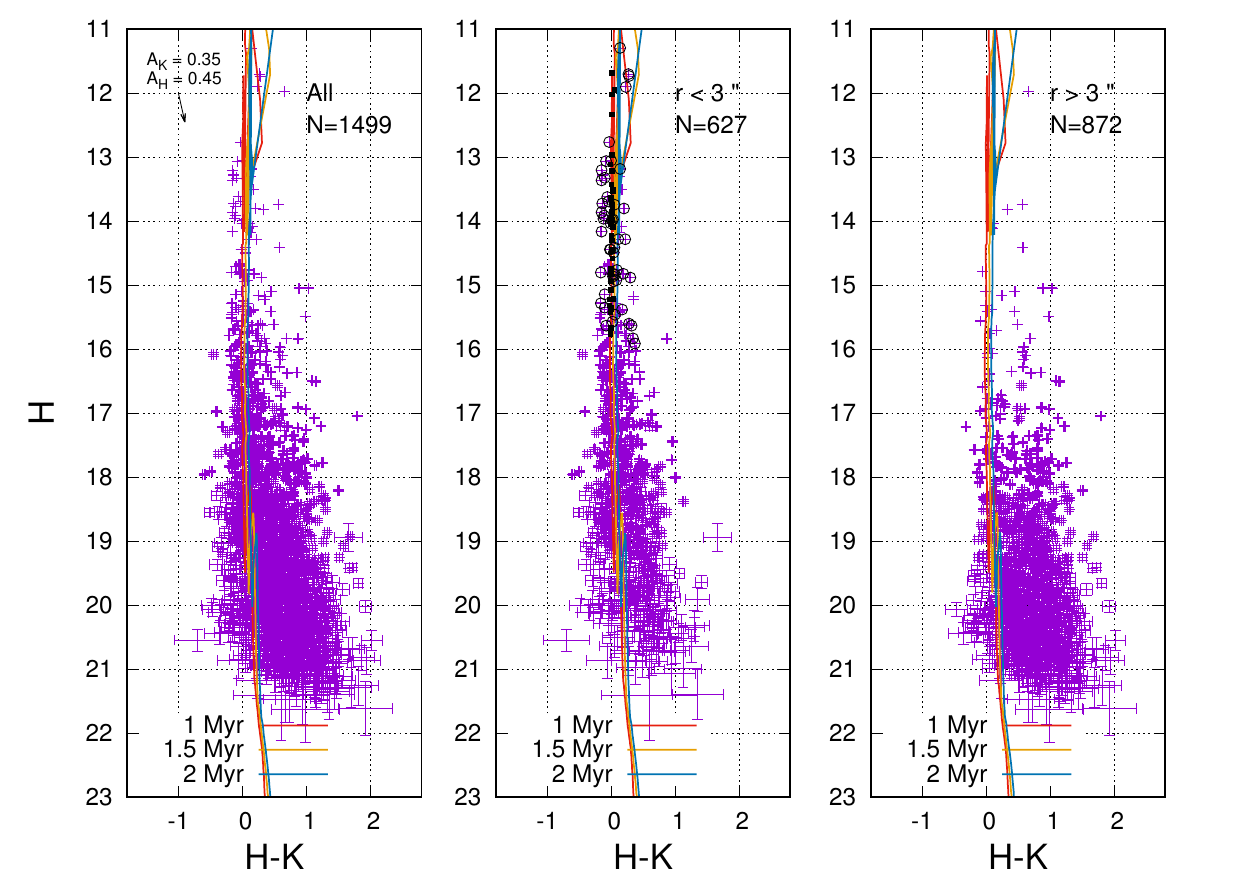}\\
    \includegraphics{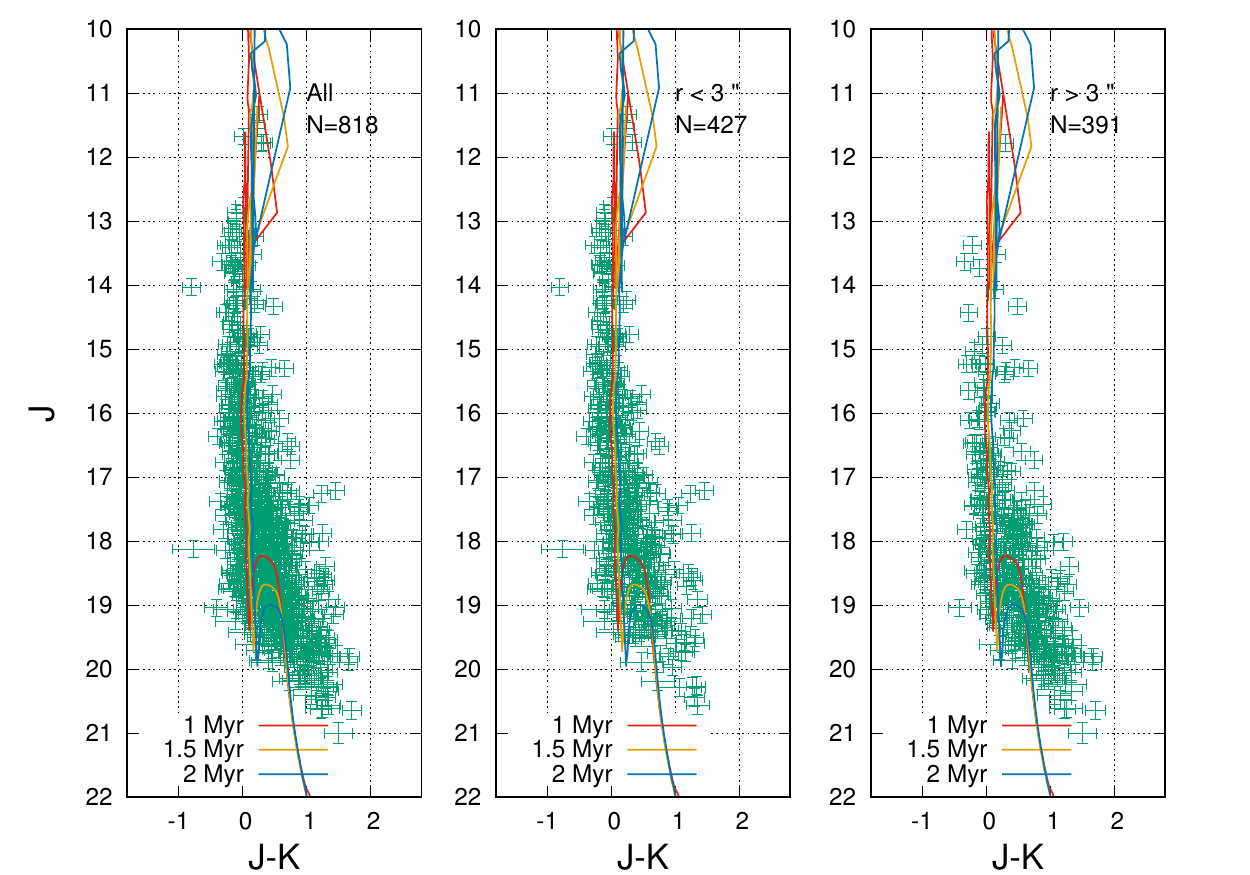}
    \caption{Top: CMD of detected sources common between the H and K data for the full FOV in left (N=1499), for central $r < 3"$ region (N=627) in middle, and for outer $r > 3"$ region (N=872) in right. Red, yellow and blue solid lines are PARSEC isochrones at DM=18.49 with $A_K=0.35$ and $E(H-K)=0.1$, at the ages of 1, 1.5, and 2 Myr, respectively. The black circles are the 49 spectroscopically-known stars from Bestenlehner et al. (2020) and the black squares represents the simulated magnitudes for these 49 sources (explained in Section \ref{sec:mass} \textit{Method II}).
    Bottom plots are same as tops but for 818 sources detected in J and K in our first epoch observation (Khorrami et al. 2017).}
    \label{fig:cmdR3}
\end{figure*}

Figure \ref{fig:cmdR3} shows the colour magnitude diagram (CMD) of 1499 detected sources common between the H and K data (top) for the full FOV in left (N=1499), for inner $r < 3"$ region (N=627) in middle, and for outer $r > 3"$ region (N=872) in right{\footnote{In the first epoch data, in J data, the PSF is not centrosymmetric at large distances from R136a1, starting typically around 3" ($\sim$ half of FOV radius). For the sake of comparison of two epochs we choose $r=3"$ for analysing detected sources in/out of this radius.}}. 
Red, yellow and blue solid lines are PARSEC isochrones at DM=18.49 with $A_K=0.35$ and $E(H-K)=0.1$, at the ages of 1, 1.5, and 2 Myr, respectively. 
We note that there is a large scatter in the CMD, especially for the lower part of the main sequence and pre-main sequence sources which are located (detected) in the outer region. 
A significant scatter in CMD and colour-colour diagrams of 30 Doradus has previously been reported in the visible and near infrared \citep{hunter95,andersen2009, demarchi2011,Cignoni2015}.
The scatter is likely due to a combination of  observational confusion (affected mainly by the visual multiple systems or variables), photometric errors, differential extinction, a possible age spread, colour excess due to warm circumstellar matter and ongoing star formation within 30 Doradus. 
Pre-main sequence stars are often associated with circumstellar disks and outflows which will introduce additional extinction for the clusters low-mass content. \cite{Brandl96} found that the extinction varies significantly from star to star within the cluster, with the range of 1-2 mag. The HST observations also reveal the presence of considerable differential extinction across the 30 Doradus region.
\cite{demarchi2011} quantified the total extinction toward massive main sequence stars younger than 3 Myr to be in the range $1.3 < A_V < 2.1$.
Figure \ref{fig:cmdR3} shows the evolutionary models (isochrones) can not be fitted for these scattered sources. We prefer to consider a (larger) error of 0.5\,mag on the extinction, to estimate the stellar mass range for each star at a given age. We estimated the stellar masses just for common sources between H and K data (1499 sources) using both their H and K magnitudes fitted to PARSEC isochrones at three different ages: 1, 1.5, and 2 Myr.

\begin{figure}
    \centering
    \includegraphics[trim=0 0 0 0,clip,width=8.2cm]{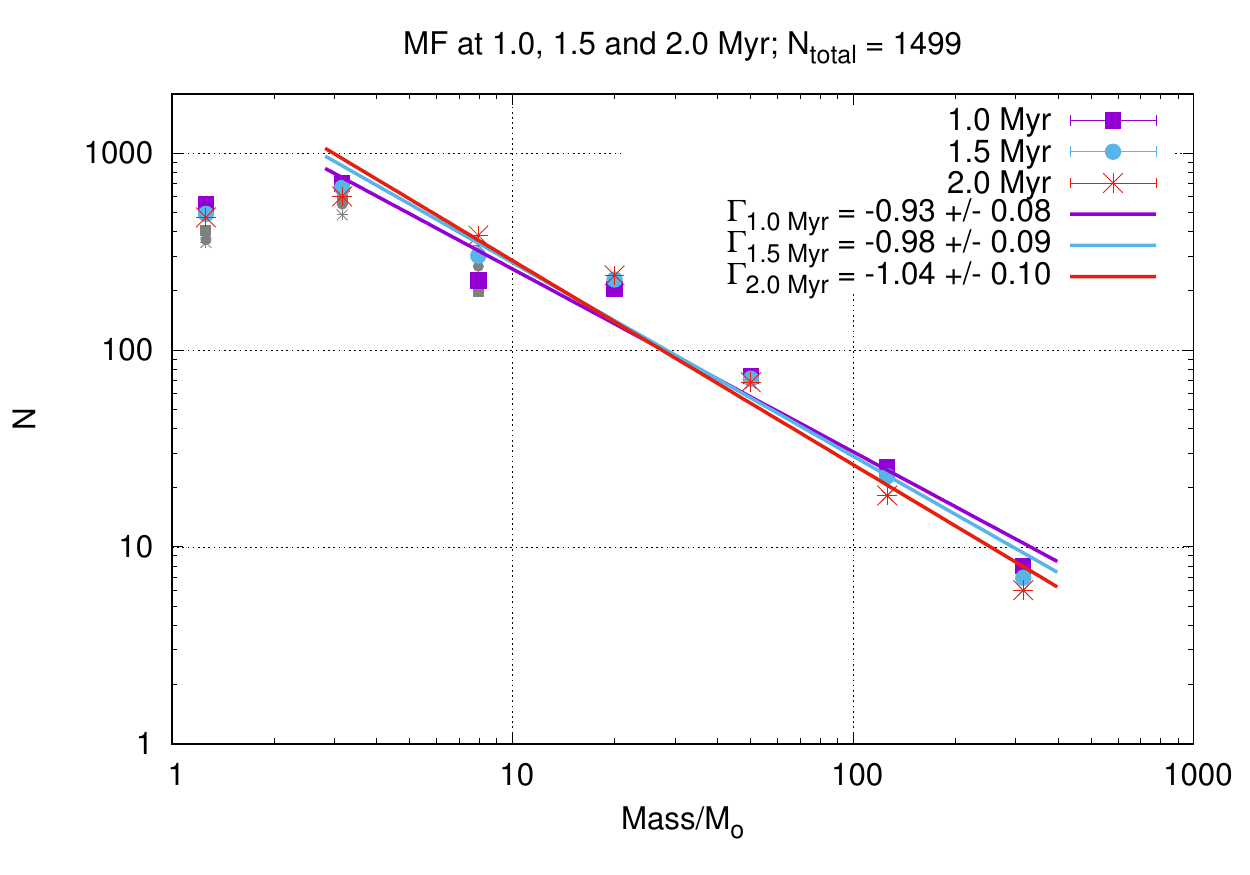}\\
        \includegraphics[trim=0 0 0 0,clip,width=8.2cm]{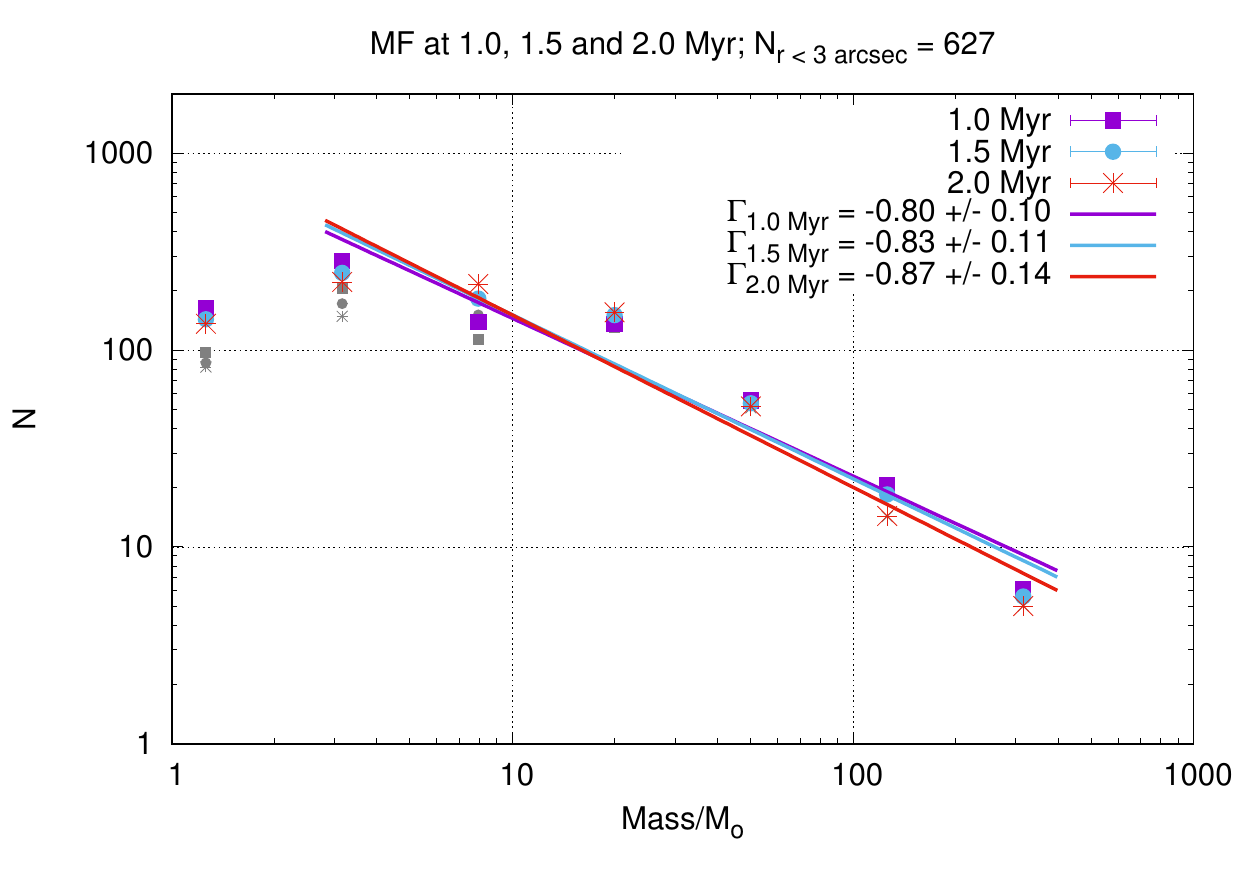}\\
    \includegraphics[trim=0 0 0 0,clip,width=8.2cm]{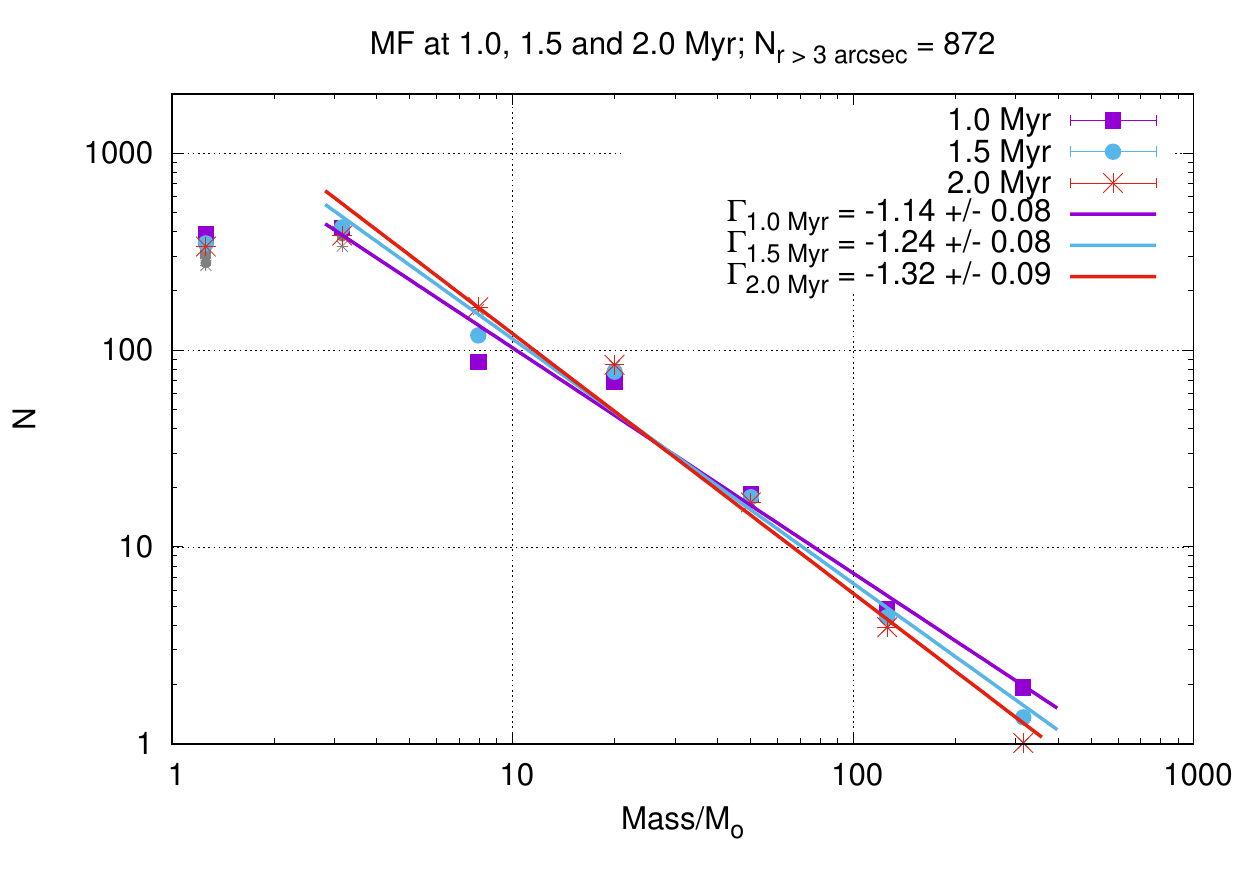}
    \caption{Mass Function at 1, 1.5 and 2 Myr, for the whole FOV (top), central r < 3" region (middle), and outer region r > 3" (bottom). Gray symbols are the values without completeness correction. See Table \ref{table:mf} for the MF slopes values.} 
    \label{fig:mf}
\end{figure}

The mass function (MF), is plotted in Figure \ref{fig:mf} considering a Gaussian distribution for each stellar mass. 
We convert the photometric/extinction uncertainty into a mass (gaussian) distribution.
This Gaussian uncertainty in the mass of each star is accounted for, when constructing the MF. 
Each star fainter than K=17 (about 13.4~M$_\odot$ at 2~Myr) is corrected for completeness, according to its brightness and its location in the FOV (see Section \ref{sec:com} for more information).
Figure \ref{fig:mf} shows the MF at three different ages (1, 1.5, and 2 Myr), for the whole FOV (top), central $r \leq 3"$ region (middle), and outer r > 3" region (bottom). The MF slope values in these regions, at different ages and mass ranges are provided in Table \ref{table:mf}, both for the completeness corrected ($\Gamma_{CC}$) and not corrected MF ($\Gamma_{NC}$).
The MF slopes are calculated for the minimum mass of 3 M$_\odot$ (about K=19.8 at 2~Myr) where the average value of completeness is above 85\% 
(Figure \ref{fig:com}) but the difference between $\Gamma_{CC}$ and $\Gamma_{NC}$ is not negligible for the inner-region of the cluster where the completeness reaches to 30\% (see Figure \ref{fig:commap}) and this affects the MF slope for the whole FOV (top values in Table \ref{table:mf}). The slopes for the mass range of $3 - 300$ M$_\odot$ for the inner region is flatter than the outer region and consequently for the whole FOV. This can be explained by completeness since the $\Gamma_{CC}$ and $\Gamma_{NC}$ are not compatible.
But for the mass range of $10 - 300$ M$_\odot$ where the data are complete (see Figure \ref{fig:com} and Figure \ref{extra:incom}) and the MFs are not affected by completeness (compare $\Gamma_{CC}$ and $\Gamma_{NC}$ in table \ref{table:mf}), the slope values in the inner region is slightly flatter than the outer region, and consequently than the whole FOV.
Comparing the stellar population in the inner region with the outer in CMD (Figure \ref{fig:cmdR3}), detected sources are compatible with the evolutionary models in the inner region (Figure \ref{fig:cmdR3}-top middle), but for the outer region detected sources differ from the isochrones. About 50\% of the high mass stars are located in the right side of the main-sequence, far from the evolutionary models. 
If, these population truly belong to R136 then they have K bandpass excess emission due to hot dust (K-excess sources).
The MF slopes in the outer region are complete for the both mass ranges and the slope values are steeper than the inner region, but might be the effect of the existence of these K-excess sources. 

The MF slopes for the massive part (M > 10 M$_\odot$) in the inner region is in line with \cite{Bestenlehner2020} ($\gamma\sim2.0\pm0.3$) and for the whole FOV is consistent with \cite{schneider2018} ($\gamma \sim 1.90^{+0.37}_{-0.26}$) for the 30 Doradus region, considering the large error bars from these studies.
\cite{massey98} found the MF slope of $-1.4 <\Gamma <-1.3$ for the mass range of 15-120~M$_\odot$ which is consistent with the slope for the whole FOV at 2 Myr.

\begin{table}
   \centering
       \caption{$\Gamma_{\rm{CC}}$ is the slope of the completeness corrected MF and $\Gamma_{\rm{NC}}$ is the not-corrected one. 
    The stellar masses are estimated using parsec isochrones with DM=18.49, A$_K=0.35$ and E(H-K)=0.1, at different ages (first column) and mass ranges (second column), for the whole FOV (top), and central r < 3" region (middle), and outer r > 3" region (bottom). N is the number of detected sources and N$_C$ is the number of stars fainter than K=17 mag that are corrected for completeness according to their position on the image and their brightness (see Section \ref{sec:com}).
    }
    \label{table:mf}
    \begin{tabular}{|c| c| c c|}
    \hline
        Age[Myr] & Mass range [M$_\odot$]&$\Gamma_{\rm{CC}}$  & $\Gamma_{\rm{NC}}$ \\
        \hline
 \multicolumn{4}{|c|}{All,  N = 1499, N$_C$ = 1226} \\   
        \hline
         1.0& 3 - 300 &$-0.93\pm0.08$ & $-0.89\pm0.09$\\
         1.5& 3 - 300 &$-0.98\pm0.09$ & $-0.94\pm0.10$\\
         2.0& 3 - 300 &$-1.04\pm0.10$ & $-0.99\pm0.11$\\
& & &\\
         1.0& 10 - 300 &$-1.17\pm0.02$ & $-1.16\pm0.03$\\
         1.5& 10 - 300 &$-1.26\pm0.01$ & $-1.25\pm0.01$\\
         2.0& 10 - 300 &$-1.34\pm0.03$ & $-1.33\pm0.03$\\
         \hline
 \multicolumn{4}{|c|}{ r < 3", N = 627 , N$_C$ = 444}\\
        \hline
         1.0& 3 - 300 &$-0.80\pm0.10$ & $-0.74\pm0.12$\\
         1.5& 3 - 300 &$-0.83\pm0.11$ & $-0.76\pm0.14$\\
         2.0& 3 - 300 &$-0.87\pm0.14$ & $-0.79\pm0.16$\\
& & &\\
         1.0& 10 - 300 &$-1.12\pm0.06$ & $-1.10\pm0.06$\\
         1.5& 10 - 300 &$-1.19\pm0.03$ & $-1.17\pm0.03$\\
         2.0& 10 - 300 &$-1.26\pm0.03$ & $-1.24\pm0.03$\\
         \hline
 \multicolumn{4}{|c|}{ r > 3" , N = 872 , N$_C$ = 782}\\
        \hline
         1.0& 3 - 300 &$-1.14\pm0.08$ & $-1.13\pm0.08$\\
         1.5& 3 - 300 &$-1.24\pm0.08$ & $-1.22\pm0.08$\\
         2.0& 3 - 300 &$-1.32\pm0.09$ & $-1.30\pm0.10$\\   
& & &\\
         1.0& 10 - 300 &$-1.31\pm0.08$ & $-1.31\pm0.08$\\
         1.5& 10 - 300 &$-1.47\pm0.06$ & $-1.47\pm0.05$\\
         2.0& 10 - 300 &$-1.60\pm0.05$ & $-1.59\pm0.04$\\
         \hline
    \end{tabular}
\end{table}

\begin{figure}
    \centering
    \includegraphics[trim=0 30 0 20,clip,width=8.2cm]{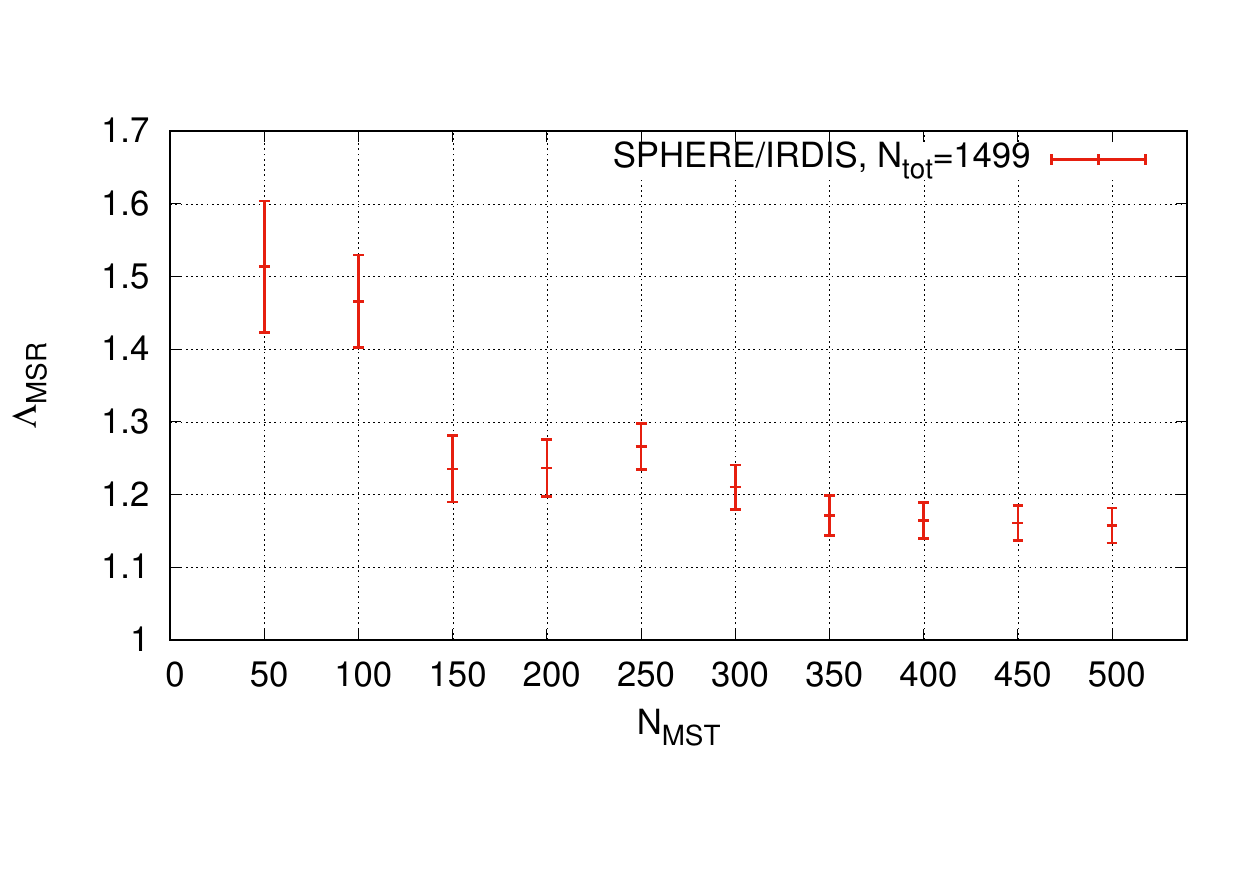}\\
    \includegraphics[trim=0 30 0 30,clip,width=8.2cm]{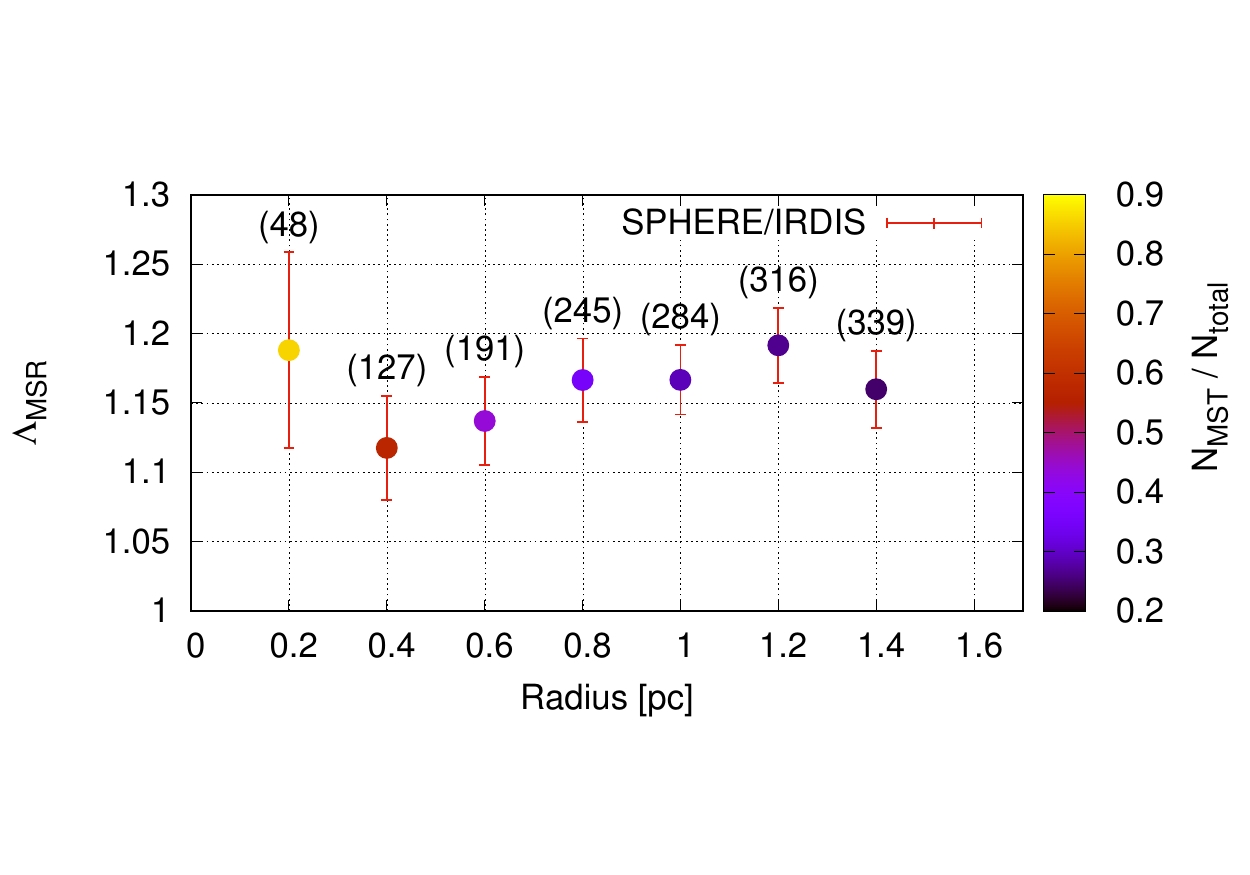}
    \caption{Top: the mass-segregation ratio ($\Lambda_{MSR}$) calculated for different $N_{MST}$ subsets. 
    Bottom: $\Lambda_{MSR}$ calculated within different radii of the cluster, centered at R136a1. The numbers in parentheses denote the total number of stars more massive than 10 M$_{\odot}$ which are chosen for $N_{MST}$. The colour shows the ratio of number of massive stars ($m > 10 M_\odot$) to the total number of stars within each radius. }
    \label{fig:mst}
\end{figure}

To check the presence of mass-segregation, we used the minimum spanning tree (MST) algorithm by \cite{allison2009} comparing the length of the MST connecting massive stars to that connecting randomly
selected stars.
Mass-segregation ratio ($\Lambda_{MSR}$, Eq. 1 in \cite{allison2009}), defined as the average random path length of the MST and that of the massive stars. $N_{MST}$ is the number of selected massive stars and randomly selected stars.
Figure \ref{fig:mst} shows $\Lambda_{MSR}$ calculated for different $N_{MST}$ subsets.
$\Lambda_{MSR}$ reaches to a constant value when $N_{MST}$ is larger than the total number of massive stars (m > 10 M$_\odot$) in the FOV which is about 339 using 1.5 Myr isochrones.
The bottom plot in Figure \ref{fig:mst} shows the $\Lambda_{MSR}$ calculated within different radii of the cluster, centered at R136a1. 
The numbers in parentheses denote the total number of stars more massive than 10 M$_{\odot}$ which are chosen for $N_{MST}$. 
The colour shows the ratio of number of massive stars ($m > 10 M_\odot$) to the total number of stars (down to 1 M$_\odot$) within each radius.
This plot shows the variation of $\Lambda_{MSR}$ locally for different sample of $N_{MST}$.
$\Lambda_{MSR}$ is less than 1.2 within the whole FOV and for $N_{MST} > 350$, and it is larger than 1.0 in all the conditions. The values of $\Lambda$ are point to a moderate but significant (above 2 sigma) degree of mass segregation.
This must be taken with caution, considering the spatial distribution of completeness (Figure \ref{extra:incom}).  The fact that, globally, we detect less low mass stars towards the centre of the cluster can have an effect on the mass segregation ratio. 
If more low mass stars are injected in the centre of the cluster, the size of the N most massive stars would still be the same, but the average size of a group of N stars within the complete sample is expected to decrease. 
The values of $\Lambda_{MSR}$ would also decrease in that case, but quantifying this is out of the scope of this work. 

Still, these MF slopes in our study are limited to the resolution of the instrument (55 mas in K) and in future, using higher angular resolution data, we may resolve binaries and low-mass stars which affects the shape of MF.

\begin{figure}
\centering
\includegraphics[trim=0 0 0 0,clip,width=8.2cm]{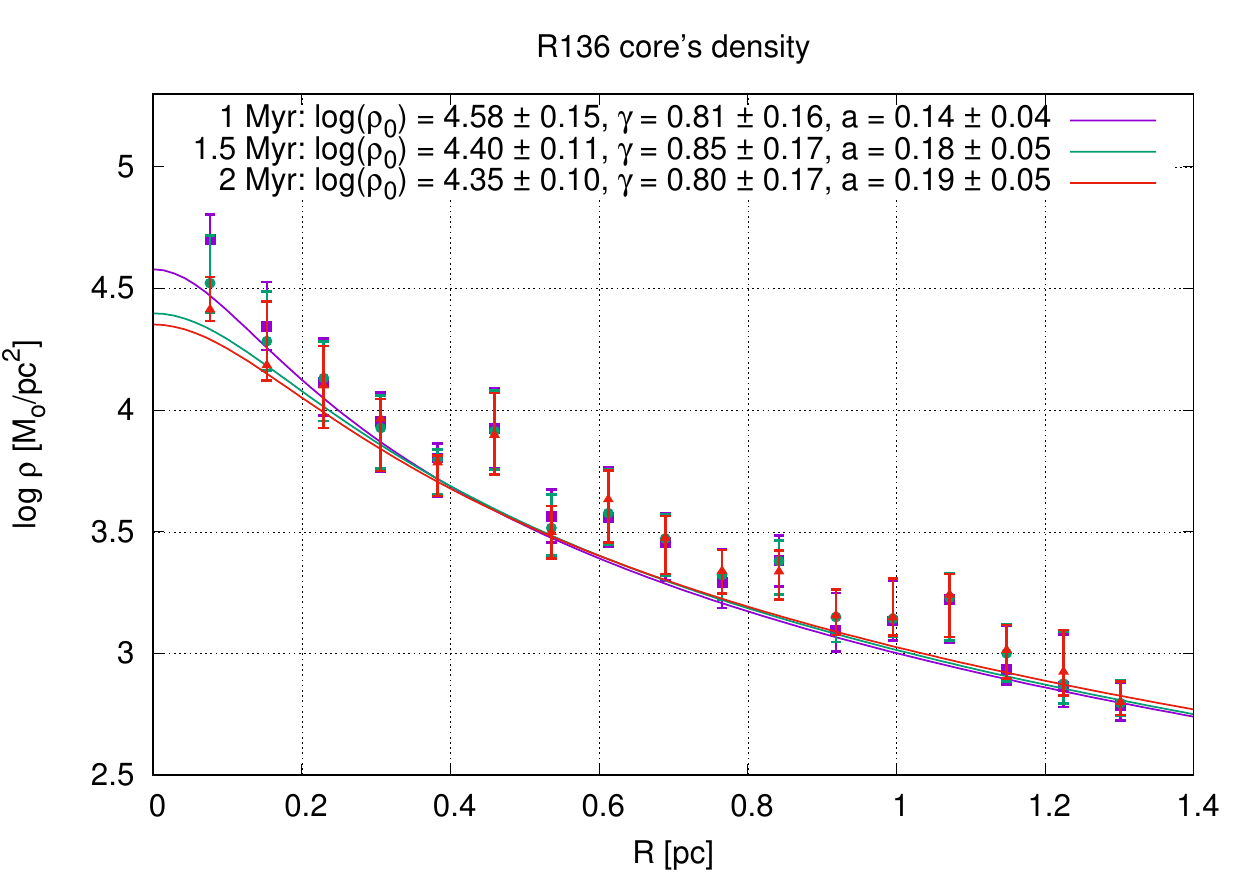}\\
\includegraphics[trim=0 10 0 30,clip,width=8.2cm]{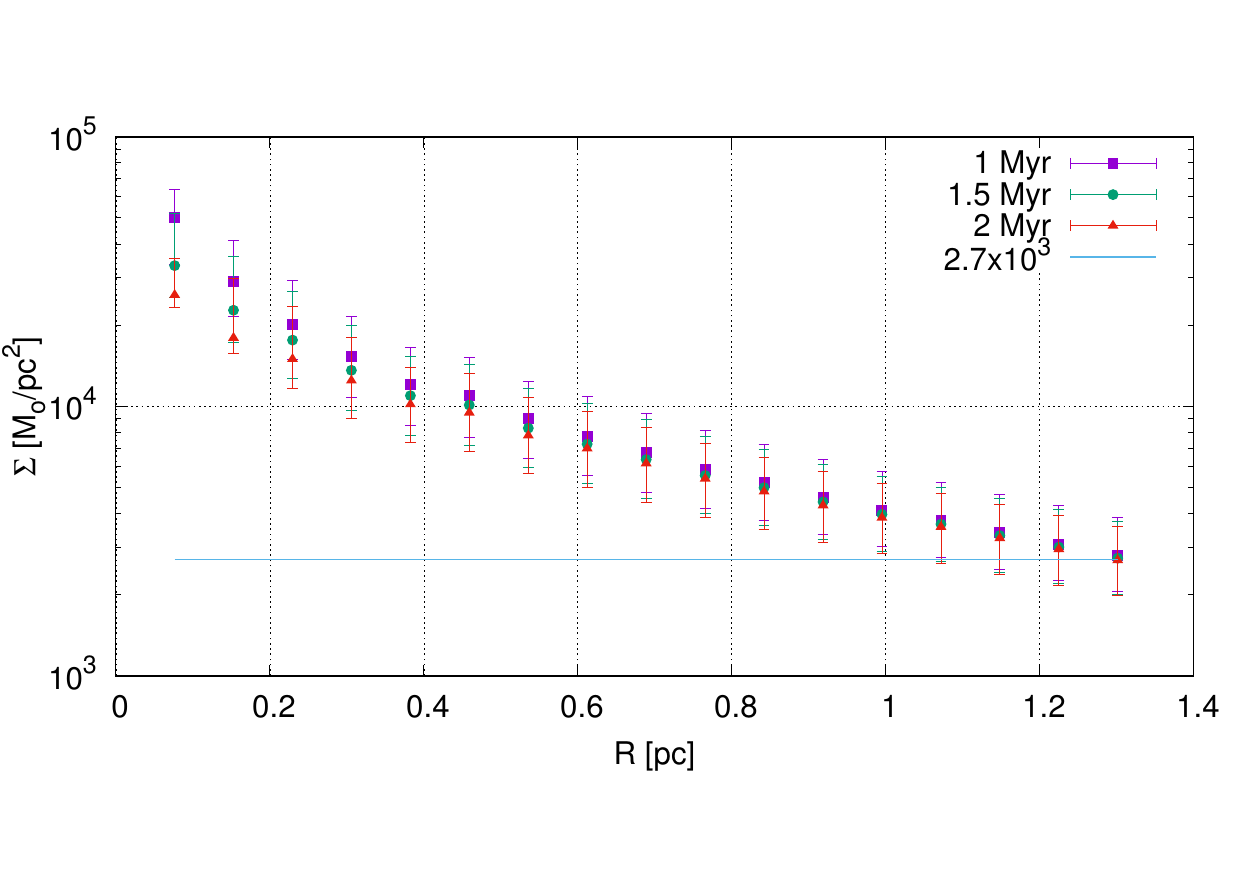}\\
\includegraphics[trim=0 20 0 35,clip,width=8.2cm]{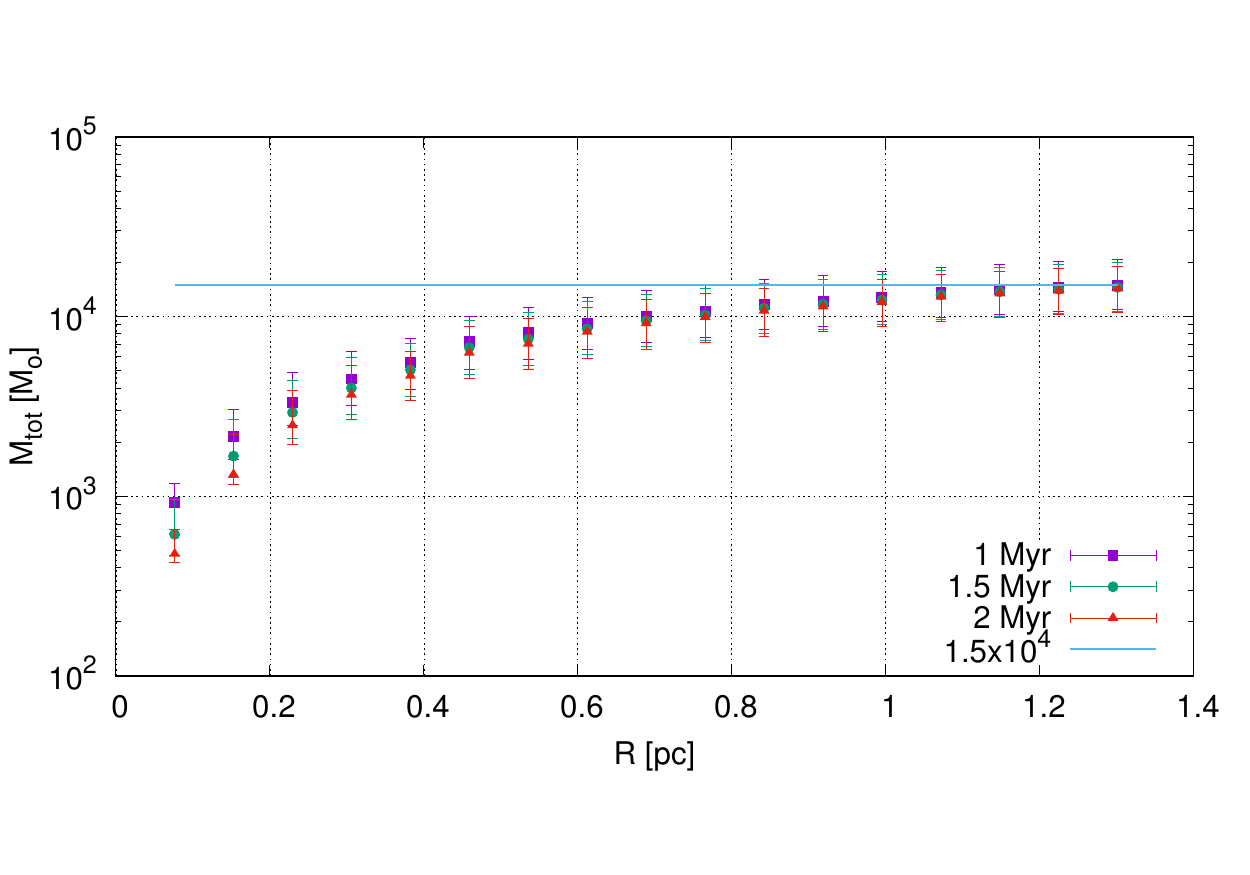}
\caption{Top: projected mass density ($\rho$) profile of R136 in IRDIS FoV centered on R136a1, using 1499 stars in H-K. 
The stellar masses are estimated at the age of 1 (purple squares), 1.5 (green circles), and 2 Myr (red triangles) with extinction values of $A_K=(0.35\pm0.5)~mag$, $E(H-K)=(0.1\pm0.1)~mag$ in H and K. Eq. \ref{eq:density} is used to fit the purple, green, and red solid lines to the data at 1, 1.5, and 2 Myr, respectively.
Center: surface densities ($\Sigma$) within a given radius. Bottom: total stellar masses ($M_{tot}$) within a given radius. Solid blue lines shows the average value of surface density ($\Sigma=2.7\times10^3 M_\odot/pc^2$) and total mass ($M_{tot}=1.5\times10^4 M_\odot$) up to $R_{max}=1.3pc$}
\label{fig:density}
\end{figure}

Using the stellar masses estimated at the age of 1, 1.5, and 2 Myr, which are corrected {\footnote{the reported values are consistent for ones derived without completeness correction. To compare these plots w/o completeness at 1.5 Myr see Figure \ref{extra:densityCC}}} for the completeness (for 1226 sources),
we plot the 2D (projected) mass density at a given radius, i.e. the mass between $r$ and $r+dr$ divided by the corresponding area ($\rho$, Figure \ref{fig:density}-top). 
We used an Elson-Fall-Freeman (EFF) profile \cite{elson1987} to fit $\rho$ in the core of R136 (Eq. \ref{eq:density}) up to the maximum radius of $R_{max}=1.32$pc  from R136a1.

\begin{equation}
\rho [M_{\odot}/pc^2] =  \frac{\rho_0}{(1+\frac{r^2}{a^2})^{\frac{\gamma+1}{2}}}
\label{eq:density}
\end{equation}

We estimated the central mass density of $\rho_0=(3.80^{+1.57}_{-1.11})\times 10^4 [M_{\odot}/pc^2]$, $\rho_0=(2.51^{+0.72}_{-0.56})\times 10^4 [M_{\odot}/pc^2]$, and $\rho_0=(2.24^{+0.58}_{-0.46})\times 10^4 [M_{\odot}/pc^2]$, at the ages of 1, 1.5 and 2 Myr, respectively.
The central density at 1 Myr is higher than the other ages because most of the massive stars are located at the central part of the cluster (see Figure \ref{fig:cmdR3}-middle) and for these stars the stellar mass estimation is higher at lower ages.

The total mass of the cluster ($M_{tot}$) and the cluster's surface density ($\Sigma$) which is the projected mass density within a given radius, i.e. all the mass between 0 and r divided by the area of the corresponding circle, are shown in Figure \ref{fig:density}.
$\Sigma$ and $M_{tot}$ become consistent at different ages for larger radii (starting at 0.3pc up to 1.3pc), where lots of pre-main sequence and low-mass stars are located/detected.
The surface density of R136 at $R=0.3$pc (and $R_{max}=1.3$pc) reaches to $1.4^{+0.7}_{0.4}\times 10^4$ [M$_\odot$/pc$^2$] (and
$2.7^{+1.1}_{0.6}\times 10^3$ [M$_\odot$/pc$^2$]), and the total mass down to $1.0\pm0.1$ M$_\odot$ is 
$4.0^{+1.7}_{-1.2} \times10^3$ M$_\odot$ (and
$1.5^{+0.5}_{-0.4} \times10^4$ M$_\odot$), respectively.

The densities and total estimated masses at 1\,Myr is slightly higher than the values reported in the first epoch (compare Figure \ref{fig:density} to Figure 13. in \cite{khorrami17}).

Fitted $\gamma$ and $a$ parameters (in Eq. \ref{eq:density}), vary from 0.8 to 0.85 and 0.14 to 0.19, respectively, depending on the selected age.

Comparing these values to the previous ones estimated for 818 stars in the 2015 data in J-K data \citep{khorrami17}, $\rho_0$ and $a$ are consistent within the given errors to their previous values of $\rho_0=(3.89^{+1.60}_{-1.14})\times 10^4 [M_{\odot}/pc^2]$ and $a=0.17\pm0.05$, but $\gamma$ is decreased by about 0.4.

\section{JHK colours}\label{sec:jhk}
Among 818 sources detected in J and K in 2015 data, and 1499 sources detected in H and K in 2018, we could detect 790 sources common in total (J, H and K). 30\% of the newly detected sources in 2018 had enough SNR to be detected in the K data in 2015 (they are brighter than the faintest star in the 2015 K catalogue) but these sources were not listed in the J since the AO correction for J data is not as good as in K. These sources were either located in the central part of the cluster where AO halo (uncorrected part of the stellar signal with the FWHM of seeing) was large leading to decreasing their SNR, or in the outer part of the cluster were the PSF were distorted (see photometric selection criteria in \cite{khorrami17}) due to anisoplanatism. 

\begin{figure}
    \centering
    \includegraphics[width=0.45\textwidth]{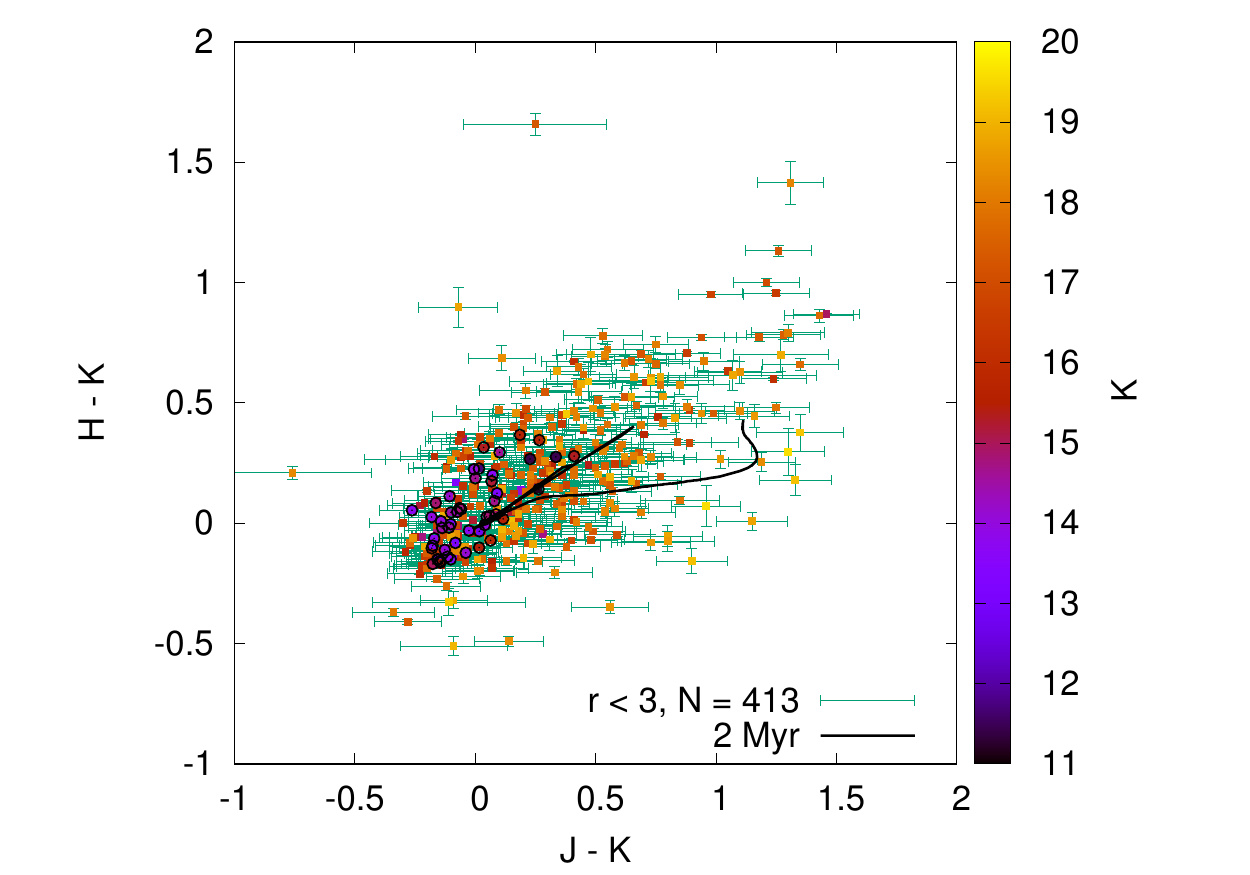}\\
    \includegraphics[width=0.45\textwidth]{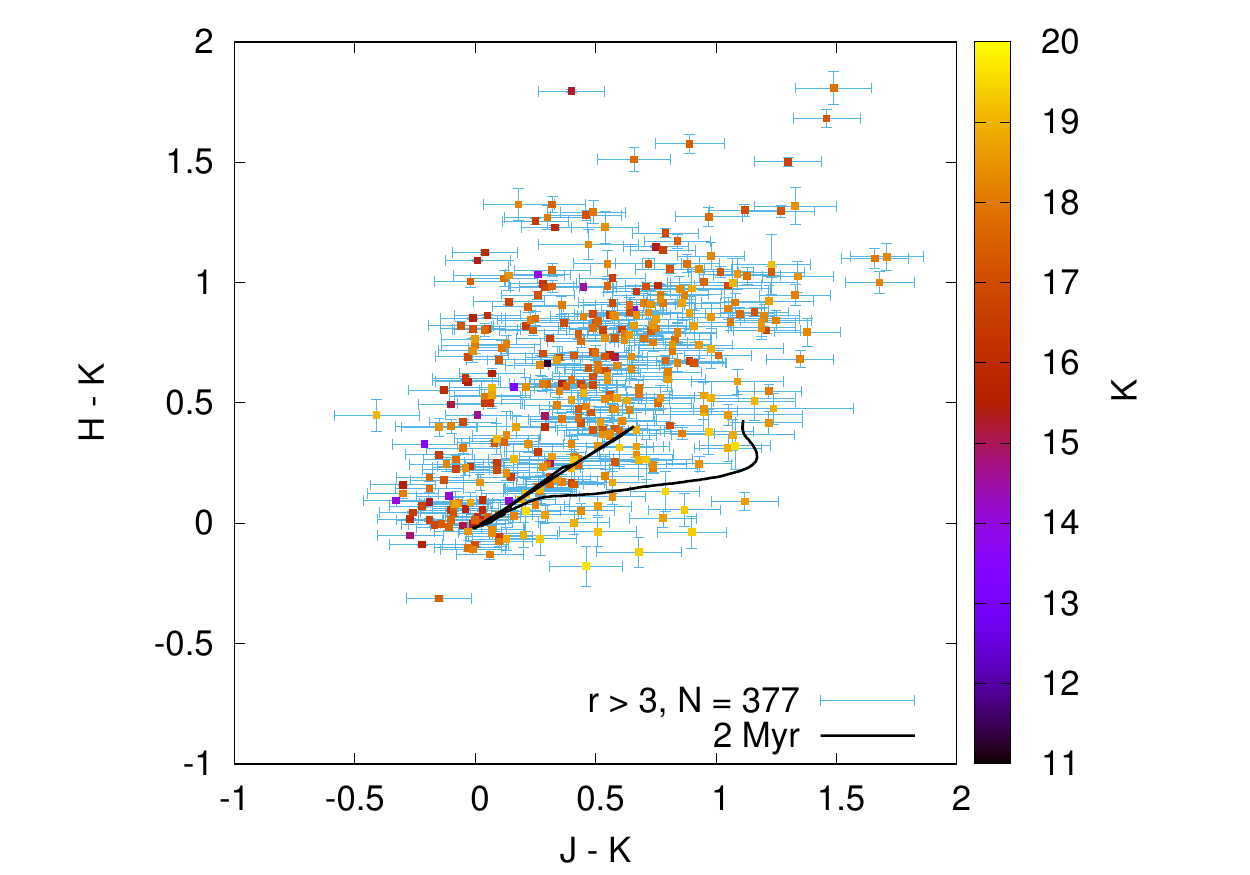}
    \caption{colour-colour diagram for the common sources between J, H and K, within two epochs J-K (2015) and H-K (2018), for the central region $r < 3"$ (top) and outer region $r > 3"$ (bottom). The colour indicates the K magnitude of detected sources and the black circles represent 49 stars studied by Bestenlehner et al 2020 listed in Table \ref{table:infostar}.
    The black solid line is the PARSEC isochrone at 2 Myr with $E(H-K)=0.1$ and $E(J-K)=0.25$. The right hand side curve represents pre-main sequence part of isochrone followed by the main sequence locus on the straight line on the centre and the post-main sequence part on the top.
     }
    \label{fig:jhk}
\end{figure}

For these 790 common sources between two epochs we have plotted colour-colour diagram in (J - K) vs (H - K) for 413 stars in the central region $r < 3"$ (Figure \ref{fig:jhk} top) and 377 sources in the outer region $r > 3"$ (Figure \ref{fig:jhk}-bottom). The black circles in Figure \ref{fig:jhk} shows 49 stars studied by \cite{Bestenlehner2020} in the centre of R136, and the black solid line is the PARSEC isochrone at 2 Myr.
These plots show that the detected sources in the inner region of R136 are more consistent with the evolutionary models, than the sources in the outer region of the cluster. 
Comparing the CMD in H-K and J-K (Figure \ref{fig:cmdR3}) for inner and outer regions, the dispersion in the colour of the detected sources in the outer region is higher than the inner region.
In the first epoch, 48\% of the detected sources in J and K were in the outer region, but in the second epoch this number increases to 58\%. Only about 30\% (70\%) of the new detected sources in the second epoch (H-K) are located in the inner (outer) region.
This can be explained by the effect of incompleteness in the core, so that even with the longer exposure time and better observing condition (e.g. SR > 70\%) central region remains too bright for low-mass MS stars to be detected.
The completeness in H and K, for the central region of R136, is less than 50\% (20\%) for the magnitude range of 19-20 (20-21), in our completeness maps (Figure \ref{extra:incom}).

\section{Conclusions}\label{sec:summary}
We presented a new photometric analysis of the core of R136 using the second epoch data from VLT/SPHERE instrument in the near-IR. 
We observed R136 in H and K filters in better atmospheric observing conditions and longer exposure time than the first epoch. This enabled us to detect twice as many sources (in H and K) as we have detected in the first epoch (in J and K), in the FOV of IRDIS ($10.8"\times 12.1"$) covering almost $2.7\times3.0$ pc of R136 core. Among 1658 and 2528 detected sources in H and K,  respectively, we found 1499 common sources between these two sets of data, where 76\%  of  these  sources  have  visual  companion closer than 0.2", which is higher than the value found in the first epoch in 2015 data \citep{khorrami17}. About 71\% of the newly detected sources are located in the outer region ($r >3"$) of the cluster.

Using the stellar parameters ($T_{eff}$, log g, and logL) of 49 stars studied spectroscopically by \cite{Bestenlehner2020} in the optical, PARSEC isochrones, Tlusty SEDs \citep{tlusty,tlustyO}, and synthetic extinction curves from \cite{drainea, draineb, drainec, LD01, WD01} we estimated the extinction at DM of 18.49 magnitude, using two methods (Section \ref{sec:mass}).
We adopt the extinctions of A$_H=0.45$ and A$_K=0.35$ magnitude, which are consistent with the two methods (within the error-bars) and \cite{demarchi14}. The colour excess of $E(H-K)=0.1$ is also consistent with \cite{tatton} and with the one used in previous studies by \cite{Campbell2010}.
Consequently, the stellar masses were calculated at the ages of 1.0, 1.5 and 2.0 Myr. 

The MF slope for 1, 1.5, and 2 Myr isochrone at the inner ($r<3"$) and outer region ($r > 3"$) of the cluster, are estimated and shown in Table \ref{table:mf}. One can check the effect of incompleteness, by comparing the MF slopes before and after the completeness correction, shown as $\Gamma_{NC}$ and $\Gamma_CC$, respectively in Table \ref{table:mf}.
The completeness-corrected MF slopes for the whole FOV ($\Gamma_{1 Myr} =-0.93 \pm 0.08$, $\Gamma_{1.5 Myr}=-0.98 \pm 0.09$), are consistent the values derived from the photometric analysis of the first epoch data \citep{khorrami17} for the mass range of (3-300) M$_\odot$ ($\Gamma_{1 Myr} =-0.90 \pm 0.13$, $\Gamma_{1.5 Myr}=-0.98 \pm 0.18$), and are closer to the Salpeter value \citep{salpeter} for the high mass range 10-300 M$_\odot$ ($\Gamma_{1.5 Myr} =-1.26 \pm 0.01$, $\Gamma_{2 Myr}=-1.34 \pm 0.03$). 
The MF slopes for the mass range of 10-300~M$_\odot$ are about 0.3 dex steeper than the mass range of 3-300~M$_\odot$, for the whole FOV and for different radii. 
The MF slopes in the inner region are shallower than the outer region for different mass ranges. 
In Figure \ref{fig:mst}, $\Lambda_{MSR}$ with a value (about 2 sigma) above 1, shows a degree of mass segregation.
Considering the spatial distribution of completeness (Figure \ref{extra:incom}), both MF slope in the core and $\Lambda_{MSR}$ would decrease, if number of low-mass stars increases in the centre where the completeness is very low.

We corrected the MF for completeness for sources fainter than 17mag (1226 stars in the whole FOV and 444/782 stars in inner/outer region). Still these values are low limits to the steepness due to incompleteness and central concentration of bright stars.

The surface density of R136 at $R=0.3$pc (and $R_{max}=1.3$pc) reaches to $1.4^{+0.7}_{0.4}\times 10^4$ [M$_\odot$/pc$^2$] (and
$2.7^{+1.1}_{0.6}\times 10^3$ [M$_\odot$/pc$^2$]), and the total mass down to $1.0\pm0.1$ M$_\odot$ is 
$4.0^{+1.7}_{-1.2} \times10^3$ M$_\odot$ (and
$1.5^{+0.5}_{-0.4} \times10^4$ M$_\odot$), respectively.
The densities and total estimated masses at 1\,Myr is slightly higher than the values reported in the first epoch (compare Figure \ref{fig:density} to Figure 13. in \cite{khorrami17}).

Comparing data with the first epoch ones, we could detect 790 sources common in total (J, H and K) and the majority (67\%) of detected sources in the outer region ($r > 3"$) are not consistent with the evolutionary models at $1-2$~Myr and with extinction similar to the average cluster value, suggesting an ongoing star formation within 30 Doradus. 
A significant scatter in the CMD (Figure \ref{fig:cmdR3}) and colour-colour diagram (Figure \ref{fig:jhk}) is originated mainly from the lower part of the main sequence and pre-main sequence sources, located (detected) at the outer region. 
The observed scatter in CMD and colour-colour diagrams of 30 Doradus has previously been reported in the visible and NIR \citep{hunter95,andersen2009, demarchi2011,Cignoni2015}. The scatter is likely due to a combination of observational confusion (affected mainly by the visual multiple systems or variables), photometric errors, differential extinction, and a possible age spread. 
In addition, pre-main sequence stars are often associated with circumstellar disks and outflows which will introduce additional extinction for the clusters low-mass content. \cite{Brandl96} found that the extinction varies significantly from star to star within the cluster, with the range of 1-2 mag. The HST observations also reveal the presence of considerable differential extinction across the 30 Doradus region.
\cite{demarchi2011} quantified the total extinction toward massive main sequence stars younger than 3 Myr to be in the range $1.3 < A_V < 2.1$.

This motivates us to observe this cluster again in J and K in future with {even} longer exposure time, so the number of common sources in three filters (J, H, and K) increases.

\section{data availability}
All data are incorporated into the article and its online supplementary material.

\section*{Acknowledgements}

The Star Form Mapper project has received funding from the European Union's Horizon 2020 research and innovation program under grant agreement No 687528. ZK acknowledges the support of a STFC Consolidated Grant (ST/K00926/1).
This work has made use of the SPHERE Data Centre, jointly operated by OSUG/IPAG (Grenoble), PYTHEAS/LAM/CeSAM (Marseille), OCA/Lagrange (Nice), Observatoire de Paris/LESIA (Paris), and Observatoire de Lyon (OSUL/CRAL). This work is supported by the French National Programms (PNPS). 
AB is funded by the European Research Council H2020-EU.1.1 ICYBOB project (Grant No. 818940). JP and RW acknowledge support by the Czech Science Foundation project no. 19-15008S and by the institutional project RVO:67985815.








\appendix

\section{Additional Table and figures}
\onecolumn

\begin{figure}
    \centering
    \includegraphics[trim=100 0 100 0,width=16cm]{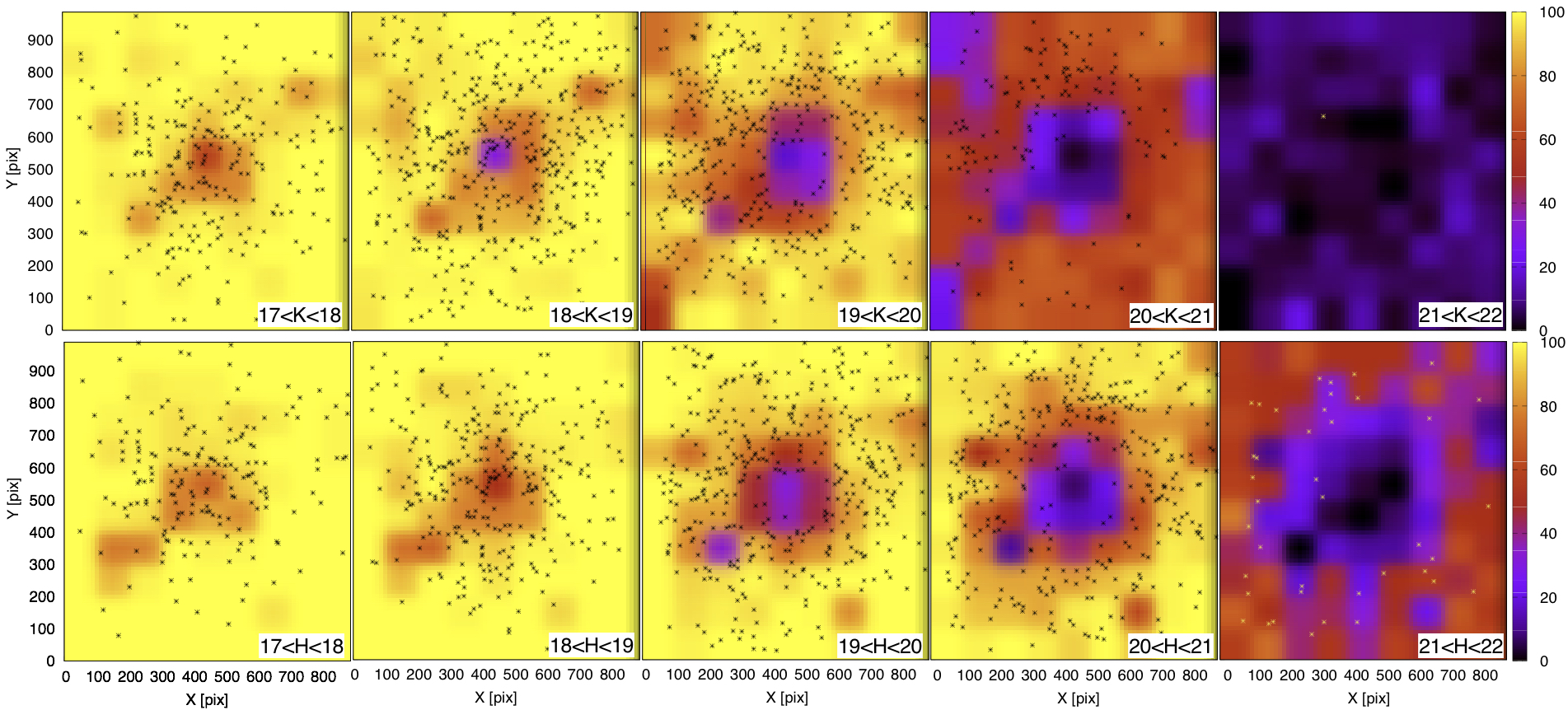}
    \caption{Completeness maps in K (top) and H (bottom), used for correcting stellar masses for plotting MF. colour range [0:100].
    Stars shows the location of the observed detected sources within the given magnitude range for each map.
    If a star with a given magnitude is located in the yellow area, it will be detected 100\% and it has difficulty to be detected if it is located in the dark-purple.}
    \label{extra:incom}
\end{figure}

\begin{figure}
    \centering
    \includegraphics[trim=50 0 50 0,clip,width=8.5cm]{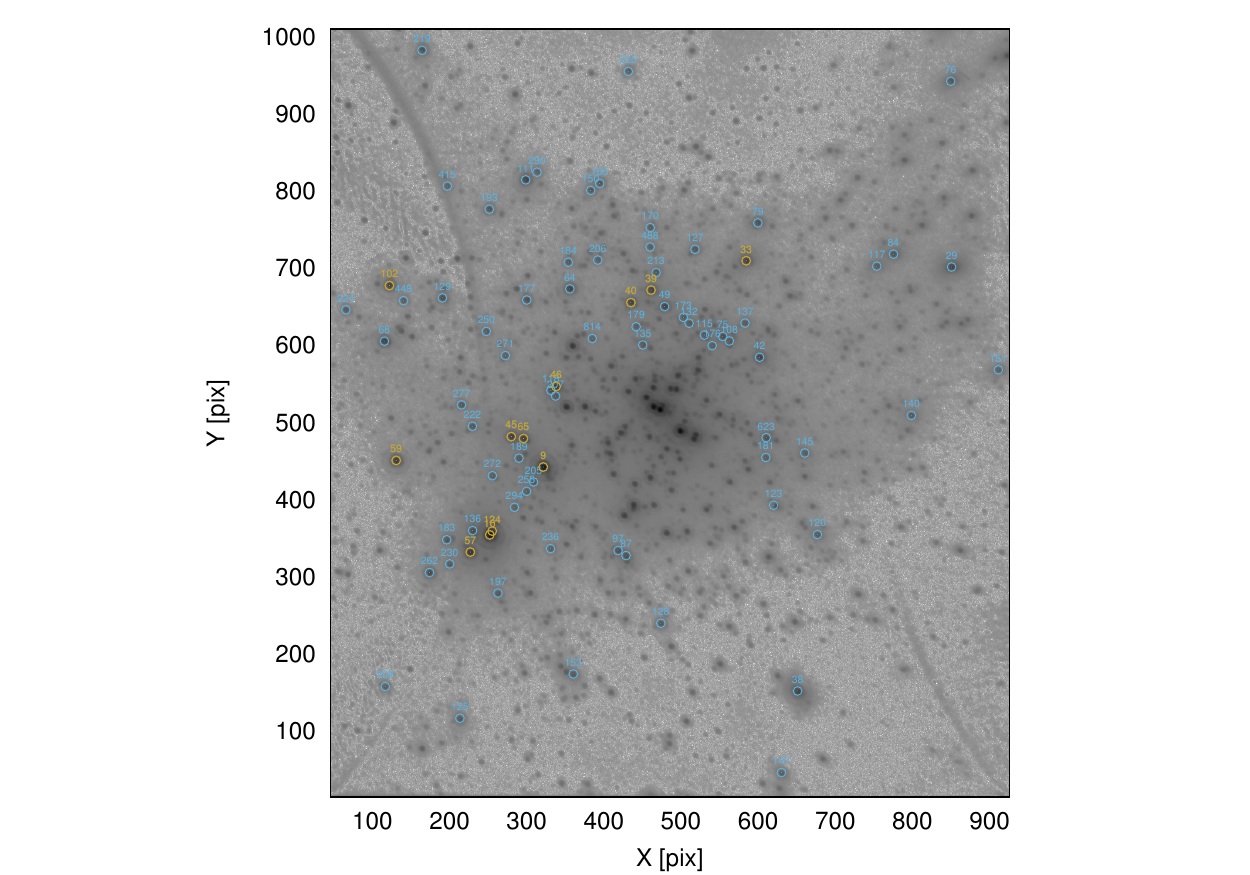}
    \includegraphics[trim=40 0 50 0,clip,width=8.5cm]{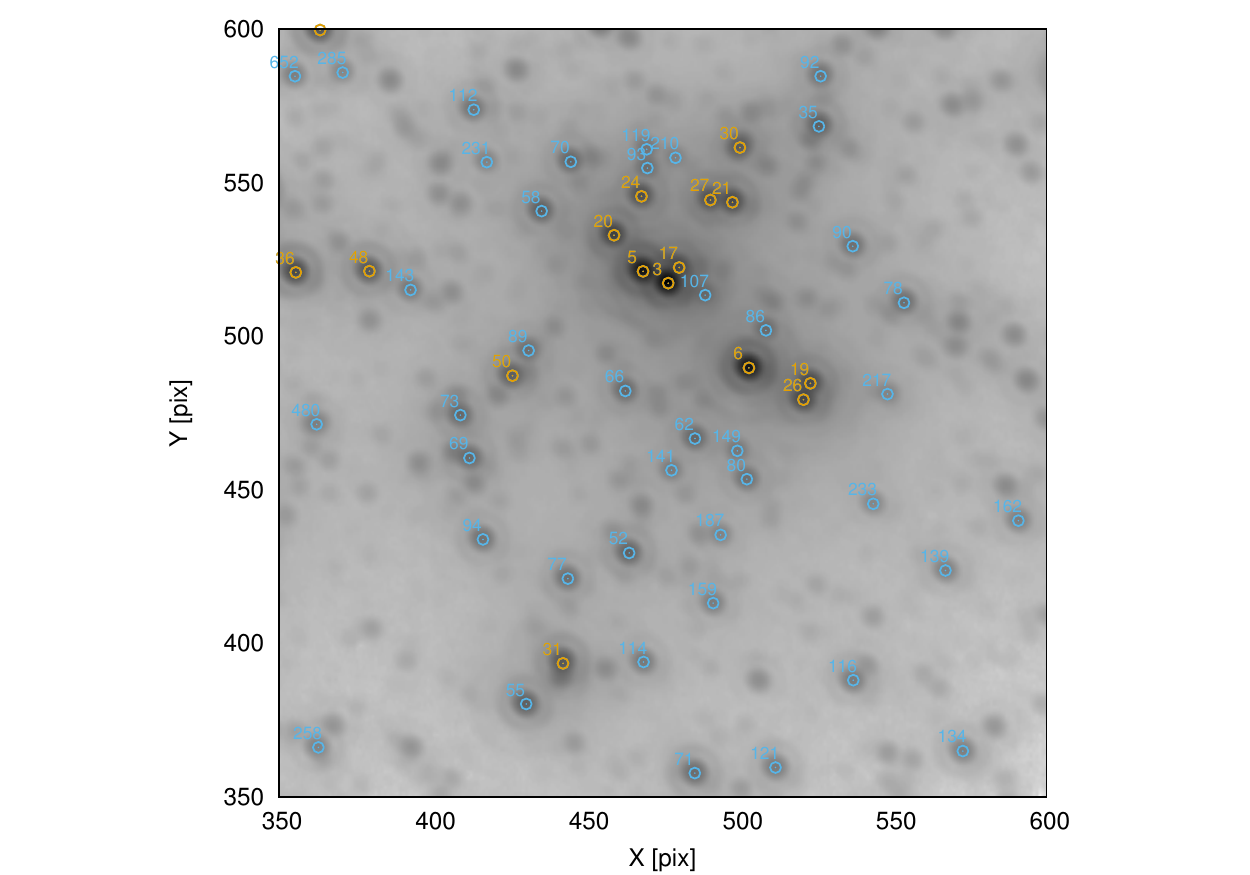}
        \caption{Stars brighter than 16~mag (14~mag) in K data shown in blue circles (yellow), with HSH95 identification written on top of each source. See Table \ref{table:bright16info} for their IRIDS (J, H, K) and HST (U, V, I) magnitudes.
        Left: whole FOV, Right: same as left but zoomed in the core to avoid confusion.
        Background image is the SPHERE/IRDIS/K in 2018.}
    \label{extra:hunter}
\end{figure}

\begin{landscape}
\begin{center}
\LTcapwidth=\textwidth
\begin{longtable}{l l l l l l l l l l l l l l c c}
    \caption{List of brightest spectroscopically known stars in the core of R136 studied in optical. 
    T$_{\rm{eff}}$, log L/L$_\odot$, and log g, and spectral types are taken from Bestenlehner et al. 2020, Table 1. Ones noted in brackets,~{\scriptsize[MH98]},~{\scriptsize[CD98]},~{\scriptsize[C10]}, ~{\scriptsize[B14]} and ~{\scriptsize[C16]} are from Massey\&Hunter (1998), Crowther \& Dessart (1998), Crowther et al. (2010), bestenlehner et al. (2014), and Crowther et al. (2016), respectively.
    HSH95 WB85, and ID$_K$ are the stars identification in Hunter et al. (1995), Weigelt \& Baier (1985), and SPHERE/IRDIS/K catalogue in 2018 (this work), respectively.
    K and H are the SPHERE IRDIS magnitudes and J$_{15}$ is the SPHERE IRDIS magnitude in our first epoch data (Khorrami et al. 2017). r is the distance from r136a1 in our K data.
    The stellar initial masses (M$_{ini}$) and ages are estimated by fitting grids of PARSEC evolutionary models (0.1 to 10 Myr) to the T$_{\rm{eff}}$ and log L/L$_\odot$. }
        \label{table:infostar}\\
 
HSH95&ID$_{K}$ & K$(\pm 0.01)$  & H$(\pm 0.01)$ &J$_{15} (\pm 0.13)$ & r&T$_{\rm{eff}}$ & log L/L$_\odot$ & log g & M$_{ini}$ & Age & Spectral \\
(WB85) & & [mag]&[mag]&[mag]&[arcsec]& [kK]&  & &[M$_\odot$]&[Myr]&type  \\
\hline
\endfirsthead
HSH95&ID$_{K}$ & K$(\pm 0.01)$  & H$(\pm 0.01)$ &J$_{15} (\pm 0.13)$ & r&T$_{\rm{eff}}$ & log L/L$_\odot$ & log g & M$_{ini}$ & Age & Spectral \\
(WB85) & & [mag]&[mag]&[mag]&[arcsec]& [kK]&  & &[M$_\odot$]&[Myr]&type  \\
\hline
\endhead
  3 (a1) &  1&11.15&11.29&11.33&0.00&$46.0\pm 2.5$&$6.79^{+0.10}_{-0.10}$& 4.0&$207^{+142}_{- 28}$&$ 2.4^{+ 0.1}_{- 1.5}$& WN5h~{\scriptsize[CD98]}\\
 10 (c)&  2&11.31&11.70&11.78&3.35&$42.0\pm2.0$~{\scriptsize[B14]}&$6.58\pm0.10$~{\scriptsize[B14]}& -&-&-&WN5h~{\scriptsize[C10]} \\
  5 (a2)&  3&11.43&11.70&11.55&0.11&$50.0\pm 2.5$&$6.75^{+0.10}_{-0.10}$& 4.0&$202^{+147}_{- 26}$&$ 2.4^{+ 0.1}_{- 1.8}$& WN5h~{\scriptsize[CD98]}\\
  6 (a3)&  4&11.45&11.73&11.77&0.47&$50.0\pm 2.5$&$6.63^{+0.10}_{-0.10}$& 4.0&$200^{+ 50}_{- 45}$&$ 0.9^{+ 1.7}_{- 0.3}$& WN5h~{\scriptsize[CD98]}\\
  9 (b)&  5&11.67&11.90&11.67&2.06&$35.0\pm 2.5$&$6.34^{+0.12}_{-0.10}$& 3.3&$102^{+ 36}_{- 23}$&$ 3.0^{+ 0.4}_{- 1.1}$& O4If/WN8~{\scriptsize[C16]}\\
 20 (a5)&  6&12.79&12.76&12.75&0.29&$47.0\pm 3.3$&$6.29^{+0.10}_{-0.09}$& 4.0&$120^{+ 19}_{- 45}$&$ 1.3^{+ 2.2}_{- 0.2}$& O2I(n)f$^*$\\
 19 (a6)&  11&13.17&13.30&13.28&0.71&$53.0\pm 3.5$&$6.27^{+0.09}_{-0.09}$& 4.1&$121^{+ 12}_{- 46}$&$ 0.7^{+ 2.8}_{- 0.6}$& O2I(n)f$^*$p\\
 36 &  8&13.05&13.05&12.86&1.49&$52.0\pm 3.4$&$6.33^{+0.12}_{-0.10}$& 4.1&$130^{+ 19}_{- 50}$&$ 0.8^{+ 2.6}_{- 0.7}$& O2If$^*$\\
 21 (a4)&  9&13.14&13.06&13.00&0.41&$48.0\pm 5.8$&$6.24^{+0.18}_{-0.18}$& 4.1&$110^{+ 39}_{- 51}$&$ 1.4^{+ 2.7}_{- 1.2}$& O3V((f$^*$))(n)\\
 24 (a7)& 13&13.35&13.20&12.95&0.36&$49.0\pm 5.5$&$6.25^{+0.18}_{-0.17}$& 4.2&$112^{+ 37}_{- 48}$&$ 1.3^{+ 2.6}_{- 1.2}$& O3III(f$^*$)~{\scriptsize[MH98]}\\
 46 & 15&13.42&13.32&13.11&1.69&$49.0\pm 6.0$&$6.16^{+0.18}_{-0.17}$& 4.2&$100^{+ 30}_{- 46}$&$ 1.2^{+ 3.2}_{- 1.1}$& O2-3III(f$^*$)\\
 47 & 17&13.51&13.36&13.17&1.72&$47.0\pm 7.0$&$6.09^{+0.22}_{-0.21}$& 4.0&$ 90^{+ 30}_{- 36}$&$ 1.4^{+ 3.0}_{- 1.3}$& O2V((f$^*$))\\
 31 & 18&13.60&13.80&13.69&1.57&$48.0\pm 5.0$&$6.01^{+0.16}_{-0.16}$& 4.0&$ 80^{+ 20}_{- 26}$&$ 1.4^{+ 3.0}_{- 1.3}$& O2V((f$^*$))\\
 48 & 19&13.68&13.62&13.41&1.19&$49.0\pm 7.2$&$6.05^{+0.21}_{-0.20}$& 4.1&$ 83^{+ 36}_{- 29}$&$ 1.5^{+ 2.9}_{- 1.4}$& O2-3III(f$^*$)\\
 45 & 20&13.69&13.74&13.37&2.39&$42.0\pm 5.0$&$5.84^{+0.17}_{-0.16}$& 3.9&$ 59^{+ 15}_{- 13}$&$ 2.5^{+ 1.9}_{- 1.0}$& O4:Vz\\
 30 & 21&13.73&13.69&13.62&0.61&$37.0\pm 3.5$&$5.68^{+0.14}_{-0.14}$& 3.9&$ 45^{+ 10}_{-  7}$&$ 3.6^{+ 0.8}_{- 0.7}$& O6.5Vz\\
 40 & 23&13.86&13.72&13.61&1.76&$45.0\pm 5.6$&$5.88^{+0.18}_{-0.18}$& 3.9&$ 65^{+ 20}_{- 15}$&$ 2.0^{+ 2.4}_{- 1.9}$& O3V\\
 65 & 27&13.94&13.97&13.67&2.21&$42.0\pm 5.2$&$5.74^{+0.17}_{-0.16}$& 4.0&$ 51^{+ 15}_{- 11}$&$ 2.8^{+ 1.0}_{- 1.9}$& O4 V~{\scriptsize[C16]}\\
 50 & 28&13.97&13.98&13.82&0.72&$42.0\pm 3.0$&$5.71^{+0.11}_{-0.11}$& 3.8&$ 50^{+  7}_{-  7}$&$ 2.7^{+ 0.7}_{- 0.7}$& O3-4 V((f$^*$))\\
 64 & 29&14.01&13.86&13.69&2.39&$40.0\pm 5.1$&$5.69^{+0.18}_{-0.17}$& 3.9&$ 47^{+ 14}_{- 10}$&$ 3.0^{+ 1.3}_{- 1.3}$& O4-5V\\
 58 & 33&14.02&13.91&13.78&0.58&$50.0\pm 5.9$&$5.94^{+0.16}_{-0.16}$& 4.1&$ 75^{+ 22}_{- 21}$&$ 1.0^{+ 3.4}_{- 0.9}$& O2-3V\\
 35 & 34&14.04&14.03&14.04&0.87&$44.0\pm5.6$&$5.74^{+0.18}_{-0.18}$& 4.0&$ 53^{+  0}_{-  0}$&$ 2.3^{+ 0.0}_{- 0.0}$& O3V\\
 55 & 36&14.06&14.28&14.15&1.77&$47.0\pm 5.0$&$5.76^{+0.15}_{-0.15}$& 3.9&$ 57^{+ 17}_{- 12}$&$ 1.5^{+ 1.3}_{- 1.4}$& O2V((f$^*$))z\\
 49 & 37&14.08&13.96&13.94&1.64&$48.0\pm12.0$&$5.89^{+0.37}_{-0.37}$& 4.2&$ 68^{+ 51}_{- 30}$&$ 1.5^{+ 3.0}_{- 1.4}$& O3V ~{\scriptsize[MH98]}\\
 52 & 39&14.17&14.28&14.10&1.09&$44.0\pm 4.8$&$5.67^{+0.16}_{-0.16}$& 4.0&$ 50^{+ 11}_{- 10}$&$ 2.2^{+ 1.3}_{- 2.1}$& O3-4Vz\\
 70 & 40&14.31&14.16&14.06&0.62&$47.0\pm 6.0$&$5.78^{+0.18}_{-0.18}$& 4.2&$ 60^{+ 20}_{- 15}$&$ 1.4^{+ 1.6}_{- 1.3}$& O5Vz\\
 62 & 41&14.36&14.41&14.28&0.63&$49.0\pm 6.2$&$5.75^{+0.17}_{-0.17}$& 4.0&$ 60^{+ 15}_{- 15}$&$ 0.8^{+ 1.8}_{- 0.7}$& O2-3V\\
 69 & 42&14.45&14.49&14.29&1.06&$41.0\pm 4.6$&$5.51^{+0.16}_{-0.16}$& 4.1&$ 40^{+ 10}_{-  8}$&$ 3.0^{+ 1.4}_{- 2.8}$& O4-5Vz\\
 66 & 43&14.45&14.44&14.31&0.46&$46.0\pm 6.6$&$5.64^{+0.21}_{-0.21}$& 4.1&$ 50^{+ 20}_{- 13}$&$ 1.7^{+ 1.8}_{- 1.6}$& O2V-III(f$^*$)\\
 73 & 44&14.46&14.44&14.24&0.98&$33.0\pm 3.6$&$5.27^{+0.14}_{-0.14}$& 4.3&$ 27^{+  4}_{-  4}$&$ 5.6^{+ 2.0}_{- 1.1}$& O9.7-B0V\\
 71 & 46&14.58&14.88&14.80&1.96&$48.0\pm 8.0$&$5.56^{+0.23}_{-0.23}$& 3.9&$ 50^{+ 15}_{- 16}$&$ 0.2^{+ 3.2}_{- 0.1}$& O2-3V((f$^*$))\\
 86 & 47&14.63&14.82&14.80&0.43&$41.0\pm 5.0$&$5.26^{+0.16}_{-0.16}$& 3.8&$ 31^{+ 10}_{-  5}$&$ 3.1^{+ 1.9}_{- 3.0}$& O5:V\\
 78 & 49&14.67&14.76&14.76&0.94&$48.0\pm 8.0$&$5.60^{+0.24}_{-0.24}$& 4.2&$ 50^{+ 20}_{- 15}$&$ 0.9^{+ 2.5}_{- 0.8}$& O4:V\\
 80 & 54&14.77&14.83&14.71&0.84&$35.0\pm 3.8$&$5.15^{+0.15}_{-0.15}$& 3.8&$ 25^{+  4}_{-  4}$&$ 5.3^{+ 1.9}_{- 2.3}$& O8V\\
 75 & 55&14.79&14.82&14.84&1.53&$39.0\pm 6.9$&$5.29^{+0.22}_{-0.22}$& 4.3&$ 31^{+ 11}_{-  8}$&$ 3.7^{+ 2.9}_{- 3.6}$& O6V\\
 90 & 59&14.83&14.86&14.88&0.75&$40.0\pm 3.7$&$5.32^{+0.13}_{-0.13}$& 4.1&$ 32^{+  7}_{-  5}$&$ 3.4^{+ 1.4}_{- 2.9}$& O4:V:\\
 94 & 60&14.84&14.92&14.69&1.26&$48.0\pm 8.2$&$5.52^{+0.23}_{-0.23}$& 4.2&$ 47^{+ 12}_{- 15}$&$ 0.3^{+ 3.1}_{- 0.2}$& O4-5Vz\\
129 & 61&14.84&14.79&14.43& 0.97 &$37.0\pm 8.2$&$4.37^{+0.26}_{-0.26}$& 4.0&-&-&- \\
 92 & 64&14.90&14.94&14.96&1.02&$39.0\pm 4.0$&$5.26^{+0.14}_{-0.14}$& 4.0&$ 30^{+  6}_{-  4}$&$ 3.9^{+ 1.5}_{- 3.2}$& O6Vz\\
112 & 67&14.96&14.80&14.61&1.04&$36.0\pm 6.0$&$5.21^{+0.19}_{-0.19}$& 4.3&$ 27^{+  8}_{-  6}$&$ 5.0^{+ 2.7}_{- 4.6}$& O7-9Vz\\
114 & 72&15.21&15.38&15.33&1.51&$44.0\pm 6.8$&$5.25^{+0.21}_{-0.21}$& 4.2&$ 35^{+  8}_{-  9}$&$ 0.7^{+ 3.7}_{- 0.6}$& O5-6V\\
143 & 75&15.24&15.14&14.93&1.03&$39.0\pm 6.0$&$5.18^{+0.20}_{-0.20}$& 4.2&$ 28^{+  9}_{-  7}$&$ 3.3^{+ 3.2}_{- 3.2}$& O8-9 V-III\\
121 & 81&15.31&15.63&15.50&1.98&$34.0\pm 4.8$&$4.86^{+0.16}_{-0.16}$& 4.2&$ 20^{+  5}_{-  3}$&$ 6.3^{+ 3.5}_{- 5.7}$& O9.5V\\
116 & 82&15.32&15.60&15.86&1.75&$34.0\pm 6.1$&$4.84^{+0.16}_{-0.16}$& 3.7&$ 20^{+  5}_{-  3}$&$ 5.9^{+ 3.5}_{- 5.8}$& O7V\\
141 & 87&15.40&15.46&15.33&0.75&$32.0\pm 6.0$&$4.79^{+0.21}_{-0.21}$& 3.6&$ 18^{+  5}_{-  2}$&$ 7.8^{+ 2.2}_{- 7.7}$& O5-6V~{\scriptsize[C16]}\\
135 & 90&15.44&15.28&15.17&1.06&$33.0\pm 4.9$&$4.89^{+0.17}_{-0.17}$& 4.0&$ 20^{+  5}_{-  3}$&$ 6.9^{+ 3.1}_{- 6.3}$& B\\
132 & 91&15.45&15.35&15.38&1.45&$39.0\pm 5.8$&$5.05^{+0.20}_{-0.20}$& 4.0&$ 26^{+  8}_{-  6}$&$ 3.1^{+ 3.5}_{- 3.0}$& O7:V\\
159 & 92&15.47&15.62&15.51&0.32&$36.0\pm 8.9$&$4.93^{+0.28}_{-0.28}$& 4.3&-&-&-\\
120 & 93&15.48&16.05&16.11&0.80&$37.0\pm 6.8$&$4.81^{+0.22}_{-0.22}$& 4.3&-&-&-\\
123 & 95&15.49&15.83&15.88&2.36&$41.0\pm 6.5$&$5.01^{+0.22}_{-0.22}$& 4.1&$ 28^{+  7}_{-  8}$&$ 0.4^{+ 5.4}_{- 0.3}$& O6V\\
134 & 99&15.54&15.91&15.90&2.21&$36.0\pm 4.8$&$4.81^{+0.17}_{-0.17}$& 4.0&$ 20^{+  4}_{-  4}$&$ 3.9^{+ 4.4}_{- 3.8}$& O7Vz\\
162 &101&15.54&15.78&15.98&0.42&$37.0\pm 13.0$&$4.87^{+0.39}_{-0.39}$& 4.3&-&-&-\\
108 &102&15.55&15.57&15.71&1.55&$43.0\pm 7.6$&$5.04^{+0.24}_{-0.24}$& 4.2&$ 28^{+  6}_{-  7}$&$ 0.7^{+ 4.6}_{- 0.6}$& OVn\\
139 &105&15.56&15.78&15.72&0.40&$38.0\pm 5.1$&$4.90^{+0.17}_{-0.17}$& 4.0&-&-&-\\
173 &121&15.71&15.63&15.70&1.51&$30.0\pm10.0$&$4.65^{+0.33}_{-0.33}$& 4.3&$ 16^{+  9}_{-  2}$&$ 9.4^{+ 0.6}_{- 9.3}$&O9+V~{\scriptsize[C16]}\\

       
\end{longtable}

\end{center}
\end{landscape}

\begin{center}
\begin{longtable}{l c c c c l c c c l}\caption{List of detected sources in the centre of R136 brighter than 16~mag in K. ID$_K$ is the stars identification in our K catalogue. K(2018) and H(2018) are the K and H magnitude of stars in the second epoch data in 2018. K(2015) and J(2015) are the K and J magnitude of stars in the first epoch data in 2015. HSH95 and WB85 are the stars identification in Hunter et al. (1995) and Weigelt \& Baier (1985), followed by their U, V, and I magnitudes in the HST/WFPC2 filters. 
Last column is V-K (F555W-K(2018)).
See Figure \ref{extra:hunter} for their positions in the image. }\label{table:bright16info}\\
ID$_K$&K&H&K&J&HSH95&U&V&I&(V-K)\\
&(2018)&(2018)&(2015)&(2015)&(WB85)&(F336W)&(F555W)&(F814W)&\\
\hline
\endfirsthead
ID$_K$&K&H&K&J&HSH95&U&V&I&(V-K)\\
&(2018)&(2018)&(2015)&(2015)&(WB85)&(F336W)&(F555W)&(F814W)&\\
\hline
\endhead
   1& 11.15& 11.29& 11.07& 11.33&   3 (a1)& 11.56& 12.84& 12.18 &  1.69\\
   2& 11.31& 11.98& 11.48& 11.78&  10 (c)& 12.52& 13.47& 12.71 &  2.16\\
   3& 11.43& 11.70& 11.32& 11.55&   5 (a2)& 11.94& 12.96& 12.48 &  1.53\\
   4& 11.45& 11.73& 11.44& 11.77&   6 (a3)& 11.86& 13.01& 12.46 &  1.56\\
   5& 11.67& 11.90& 11.66& 11.67&   9 (b)& 12.29& 13.32& 12.76 &  1.65\\
   6& 12.79& 12.76& 12.73& 12.75&  20 (a5)& 12.77& 13.93& 13.54 &  1.14\\
   7& 13.04& 13.18& 13.05& 13.24&  26& 12.89& 14.19& 13.64 &  1.15\\
   8& 13.05& 13.05& 12.96& 12.86&  36& 13.36& 14.49& 13.98 &  1.44\\
   9& 13.14& 13.06& 13.09& 13.00&  21 (a4)& 12.81& 13.96& 13.63 &  0.82\\
  10& 13.17& 13.73& 13.36& 13.52&  57& 14.03& 14.87& 14.23 &  1.70\\
  11& 13.18& 13.30& 13.19& 13.28&  19 (a6)& 12.86& 13.92& 13.72 &  0.74\\
  12& 13.33& 13.50& 13.45& 13.37&  17& 13.00& 13.78& 13.89 &  0.45\\
  13& 13.35& 13.20& 13.05& 12.95&  24 (a7)& 12.84& 14.06& 13.66 &  0.71\\
  14& 13.41& 13.27& 13.27& 13.18&  27 (a8)& 12.93& 14.22& 13.86 &  0.81\\
  15& 13.42& 13.32& 13.29& 13.11&  46& 13.64& 14.73& 14.26 &  1.31\\
  16& 13.49& 13.82& 13.58& 13.37&  59& 13.81& 14.88& 14.35 &  1.39\\
  17& 13.51& 13.36& 13.32& 13.17&  47& 13.64& 14.75& 14.31 &  1.24\\
  18& 13.60& 13.80& 13.62& 13.69&  31& 13.14& 14.41& 14.08 &  0.81\\
  19& 13.68& 13.62& 13.58& 13.41&  48& 13.61& 14.78& 14.37 &  1.10\\
  20& 13.69& 13.74& 13.63& 13.37&  45& 13.51& 14.68& 14.31 &  0.99\\
  21& 13.73& 13.69& 13.65& 13.62&  30& 13.09& 14.32& 14.02 &  0.59\\
  22& 13.83& 14.41& 13.84& 14.33& 124&999.99& 15.97& 15.46 &  2.14\\
  23& 13.86& 13.72& 13.72& 13.61&  40& 13.32& 14.60& 14.35 &  0.74\\
  24& 13.87& 13.77& 13.78& 13.70&  39& 13.31& 14.59& 14.31 &  0.72\\
  25& 13.92& 14.08& 13.99& 14.23&  33& 13.07& 14.43& 14.19 &  0.51\\
  26& 13.93& 14.04& 13.84& 13.73& 102& 14.75& 15.70& 15.10 &  1.77\\
  27& 13.94& 13.97& 13.85& 13.67&  65& 14.03& 15.18& 14.49 &  1.24\\
  28& 13.97& 13.98& 13.96& 13.82&  50& 13.62& 14.81& 14.48 &  0.84\\
  29& 14.01& 13.86& 13.84& 13.69&  64& 14.01& 15.12& 14.68 &  1.11\\
  30& 14.01& 14.10& 13.93& 14.07& 111& 14.94& 15.77& 15.20 &  1.76\\
  31& 14.01& 14.11& 13.96& 13.63&  68& 14.13& 15.22& 14.76 &  1.21\\
  32& 14.01& 15.05& 14.68& 14.94&  38& 13.28& 14.57& 14.11 &  0.56\\
  33& 14.02& 13.91& 13.91& 13.78&  58& 13.65& 14.88& 14.55 &  0.86\\
  34& 14.04& 14.03& 14.00& 14.04&  35& 13.35& 14.48& 14.27 &  0.44\\
  35& 14.05& 14.18& 14.11& 14.30&  42& 13.35& 14.63& 14.58 &  0.58\\
  36& 14.06& 14.28& 14.15& 14.15&  55& 13.62& 14.86& 14.53 &  0.80\\
  37& 14.08& 13.96& 13.98& 13.94&  49& 13.55& 14.80& 14.53 &  0.72\\
  38& 14.16& 15.05& 14.64& 15.30&  29& 12.89& 14.30& 14.03 &  0.14\\
  39& 14.17& 14.28& 14.20& 14.10&  52& 13.52& 14.82& 14.55 &  0.65\\
  40& 14.31& 14.16& 14.20& 14.06&  70& 13.93& 15.23& 14.80 &  0.92\\
  41& 14.36& 14.41& 14.35& 14.28&  62& 13.72& 15.02& 14.75 &  0.66\\
  42& 14.45& 14.49& 14.39& 14.29&  69& 14.02& 15.22& 15.03 &  0.77\\
  43& 14.45& 14.44& 14.42& 14.31&  66& 13.95& 15.19& 14.88 &  0.74\\
  44& 14.46& 14.44& 14.38& 14.24&  73& 14.16& 15.27& 15.01 &  0.81\\
  45& 14.50& 15.48& 14.84& 15.29& 262& 16.08& 16.92& 16.18 &  2.42\\
  46& 14.58& 14.88& 14.70& 14.80&  71& 14.01& 15.23& 15.00 &  0.65\\
  47& 14.63& 14.82& 14.80& 14.80&  86& 14.23& 15.49& 15.19 &  0.86\\
  48& 14.65& 15.10& 14.79& 14.80& 136& 15.14& 16.06& 15.58 &  1.41\\
  49& 14.67& 14.76& 14.68& 14.76&  78& 14.11& 15.34& 15.05 &  0.67\\
  51& 14.72& 14.68& 14.49& 14.77& 107& 14.58& 15.74& 15.34 &  1.02\\
  52& 14.72& 14.69& 14.72& 14.56&  77& 13.97& 15.30& 14.99 &  0.58\\
  53& 14.76& 14.70& 14.64& 14.42& 118& 14.79& 15.89& 15.45 &  1.13\\
  54& 14.77& 14.83& 14.77& 14.71&  80& 14.13& 15.35& 15.12 &  0.58\\
  55& 14.79& 14.82& 14.78& 14.84&  75& 13.89& 15.28& 15.06 &  0.49\\
  56& 14.81& 14.70& 14.61& 14.49& 177& 15.35& 16.31& 15.76 &  1.50\\
  57& 14.81& 14.80& 14.69& 14.62&  89& 14.39& 15.52& 15.21 &  0.71\\
  58& 14.83& 15.18& 14.97& 14.92&  87& 14.32& 15.50& 15.28 &  0.67\\
  59& 14.83& 14.86& 14.83& 14.88&  90& 14.30& 15.55& 15.29 &  0.72\\
  60& 14.84& 14.92& 14.85& 14.69&  94& 14.32& 15.60& 15.30 &  0.76\\
  61& 14.84& 14.79& 14.70& 14.43& 129& 14.97& 16.03& 15.65 &  1.19\\
  62& 14.86& 14.82& 14.78& 14.81&  93& 14.34& 15.60& 15.24 &  0.74\\
  63& 14.87& 15.22& 15.02& 14.96&  97& 14.43& 15.61& 15.36 &  0.74\\
  64& 14.90& 14.94& 14.87& 14.96&  92& 14.28& 15.59& 15.38 &  0.69\\
  65& 14.91& 15.15& 15.03& 15.34&  79& 13.98& 15.35& 15.09 &  0.44\\
  66& 14.96& 15.41& 15.10& 15.39& 219& 15.81& 16.68& 16.03 &  1.72\\
  67& 14.96& 14.80& 14.78& 14.61& 112& 14.63& 15.77& 15.45 &  0.81\\
  68& 14.97& 15.84& 15.75& 17.21& 623& 18.01& 18.41& 17.45 &  3.44\\
  69& 15.10& 15.09& 14.98& 14.93& 193& 15.66& 16.51& 16.03 &  1.41\\
  70& 15.12& 15.61& 15.32& 15.22& 183& 15.33& 16.35& 15.85 &  1.23\\
  71& 15.14& 15.83& 15.50& 16.08&  84& 14.20& 15.48& 15.20 &  0.34\\
  72& 15.21& 15.38& 15.26& 15.33& 114& 14.51& 15.82& 15.59 &  0.61\\
  73& 15.21& 15.28& 15.16& 14.93& 222& 15.76& 16.70& 16.21 &  1.49\\
  74& 15.23& 15.25& 15.18& 15.17& 119& 15.00& 15.92& 15.88 &  0.69\\
  75& 15.24& 15.14& 15.11& 14.93& 143& 14.89& 16.10& 15.76 &  0.86\\
  77& 15.28& 15.33& 15.27& 15.42& 127& 14.80& 16.02& 15.73 &  0.74\\
  78& 15.29& 15.42& 15.22& 15.33& 149& 14.94& 16.14& 15.85 &  0.85\\
  79& 15.30& 15.38& 15.22& 15.03& 277& 16.07& 16.97& 16.41 &  1.67\\
  80& 15.31& 15.31& 15.25& 15.24& 150& 14.97& 16.16& 15.85 &  0.85\\
  81& 15.31& 15.63& 15.47& 15.50& 121& 14.81& 15.96& 15.70 &  0.65\\
  82& 15.32& 15.60& 15.45& 15.86& 116& 14.64& 15.85& 15.60 &  0.53\\
  83& 15.35& 16.50& 15.98& 16.73&  76& 13.92& 15.29& 15.11 & -0.06\\
  84& 15.38& 15.96& 15.63& 15.60& 197& 15.49& 16.53& 16.04 &  1.15\\
  85& 15.38& 15.62& 15.23& 15.70& 213& 15.70& 16.65& 16.38 &  1.27\\
  86& 15.40& 15.41& 15.33& 15.35& 165& 15.14& 16.23& 15.91 &  0.83\\
  87& 15.40& 15.46& 15.40& 15.33& 141& 14.74& 16.09& 15.80 &  0.69\\
  88& 15.40& 16.49& 15.96& 15.97& 125& 14.84& 15.98& 15.72 &  0.58\\
  89& 15.43& 15.38& 15.33& 15.08& 207& 15.72& 16.62& 16.28 &  1.19\\
  90& 15.44& 15.28& 15.32& 15.17& 135& 14.89& 16.05& 15.79 &  0.61\\
  91& 15.45& 15.35& 15.36& 15.38& 132& 14.77& 16.04& 15.81 &  0.59\\
  92& 15.47& 15.62& 15.51& 15.51& 159& 15.07& 16.21& 15.90 &  0.74\\
  93& 15.48& 16.05& 15.75& 16.11& 120& 14.76& 15.95& 15.69 &  0.47\\
  94& 15.48& 16.10& 15.78& 15.85& 128& 14.91& 16.02& 15.76 &  0.54\\
  95& 15.49& 15.83& 15.61& 15.88& 123& 14.76& 15.96& 15.70 &  0.47\\
  96& 15.49& 15.58& 15.47& 15.31& 205& 15.49& 16.59& 16.23 &  1.10\\
  97& 15.50& 15.58& 15.42& 15.21& 189& 15.35& 16.46& 16.13 &  0.96\\
  98& 15.51& 16.36& 15.98& 15.97& 153& 14.99& 16.17& 15.86 &  0.66\\
  99& 15.54& 15.91& 15.71& 15.90& 134& 14.83& 16.05& 15.82 &  0.51\\
 100& 15.54& 15.75& 15.46& 15.77& 480& 17.40& 17.91& 17.01 &  2.37\\
 101& 15.54& 15.79& 15.66& 15.98& 162& 15.04& 16.23& 15.92 &  0.69\\
 102& 15.55& 15.57& 15.59& 15.71& 108& 14.65& 15.74& 15.49 &  0.19\\
 103& 15.55& 15.71& 15.53& 15.23& 224& 15.56& 16.71& 16.30 &  1.16\\
 104& 15.55& 15.54& 15.51& 15.56& 170& 15.09& 16.27& 15.97 &  0.72\\
 105& 15.56& 15.78& 15.64& 15.72& 139& 14.80& 16.08& 15.86 &  0.52\\
 106& 15.56& 15.65& 15.58& 15.76& 137& 14.78& 16.06& 15.80 &  0.50\\
 107& 15.57& 16.16& 15.79& 15.75& 230& 15.75& 16.76& 16.35 &  1.19\\
 108& 15.58& 15.42& 15.43& 15.24& 184& 15.30& 16.42& 16.11 &  0.84\\
 109& 15.58& 15.88& 15.54& 15.95& 652& 18.00& 18.50& 18.05 &  2.92\\
 111& 15.62& 15.45& 15.46& 15.53& 115& 14.70& 15.83& 15.65 &  0.21\\
 112& 15.63& 15.86& 15.67& 15.55& 255& 15.94& 16.91& 16.47 &  1.28\\
 113& 15.64& 16.02& 15.83& 16.28& 145& 14.93& 16.12& 15.88 &  0.48\\
 115& 15.65& 15.56& 15.45& 15.23& 250& 16.03& 16.89& 16.49 &  1.24\\
 116& 15.68& 16.37& 16.04& 16.60& 140& 14.90& 16.08& 15.81 &  0.40\\
 117& 15.68& 15.96& 15.67& 15.61& 272& 16.17& 16.96& 16.49 &  1.28\\
 118& 15.69& 15.51& 15.52& 15.38& 179& 15.13& 16.32& 16.10 &  0.63\\
 119& 15.70& 15.76& 15.56& 15.52& 449& 17.18& 17.81& 17.15 &  2.11\\
 120& 15.70& 16.28& 16.04& 16.40& 117& 14.60& 15.86& 15.71 &  0.16\\
 121& 15.71& 15.63& 15.63& 15.70& 173& 15.00& 16.28& 16.10 &  0.57\\
 122& 15.75& 16.03& 15.88& 16.10& 181& 15.20& 16.34& 16.05 &  0.59\\
 123& 15.77& 16.05& 15.85& 15.83& 258& 15.86& 16.92& 16.42 &  1.15\\
 125& 15.79& 15.67& 15.64& 15.35& 271& 15.94& 16.96& 16.55 &  1.17\\
 126& 15.81& 15.77& 15.80& 15.79& 176& 15.02& 16.30& 16.07 &  0.49\\
 127& 15.82& 16.19& 15.74& 16.44& 488& 17.17& 17.93& 17.26 &  2.11\\
 128& 15.82& 15.70& 15.69& 15.58& 231& 15.68& 16.77& 16.42 &  0.95\\
 129& 15.83& 15.68& 15.68& 15.55& 206& 15.43& 16.60& 16.34 &  0.77\\
 130& 15.86& 15.95& 15.90& 15.83& 187& 15.12& 16.44& 16.20 &  0.58\\
 132& 15.88& 16.25& 15.98& 15.92& 236& 15.72& 16.82& 16.47 &  0.94\\
 133& 15.91& 16.05& 15.95& 16.06& 217& 15.47& 16.67& 16.39 &  0.76\\
 134& 15.91& 16.36& 15.76& 16.52& 814& 19.33& 19.01& 18.21 &  3.10\\
 135& 15.93& 15.98& 15.87& 15.90& 290& 16.11& 17.04& 16.61 &  1.11\\
 136& 15.95& 15.89& 15.88& 15.81& 210& 15.41& 16.64& 16.42 &  0.69\\
 137& 15.95& 16.35& 16.08& 16.37& 225& 15.66& 16.72& 16.38 &  0.77\\
 138& 15.95& 16.94& 16.45& 17.21& 151& 15.04& 16.16& 15.91 &  0.21\\
 139& 15.95& 16.23& 16.03& 15.86& 294& 16.00& 17.06& 16.68 &  1.11\\
 140& 15.95& 17.08& 16.49& 16.53& 209& 15.70& 16.64& 16.27 &  0.69\\
 141& 15.97& 16.17& 16.02& 16.17& 233& 15.71& 16.78& 16.52 &  0.81\\
 142& 15.97& 16.06& 15.87& 15.90& 415& 16.81& 17.64& 17.08 &  1.67\\
 143& 15.97& 17.20& 16.57& 16.90& 147& 14.92& 16.13& 15.91 &  0.16\\
 144& 16.00& 15.78& 15.81& 15.58& 285& 15.93& 17.01& 16.67 &  1.01
\end{longtable}
\end{center}

\newpage

\begin{figure}
    \centering
    \includegraphics{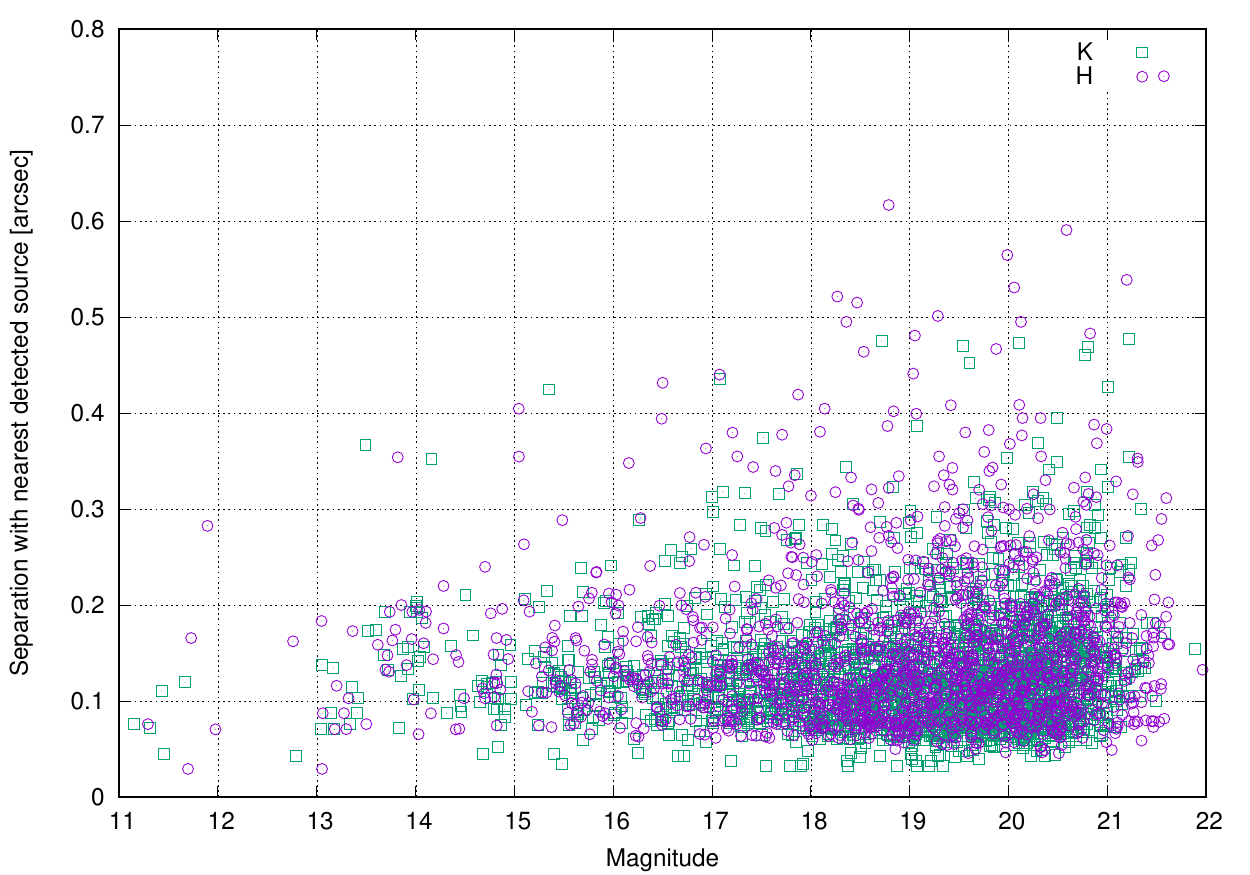}
    \caption{Separation from the nearest neighbour for each detected sources in H (purple circles) and K (green squares) versus their magnitudes.}
    \label{extra:sep}
\end{figure}

\begin{figure}
    \centering
    \includegraphics{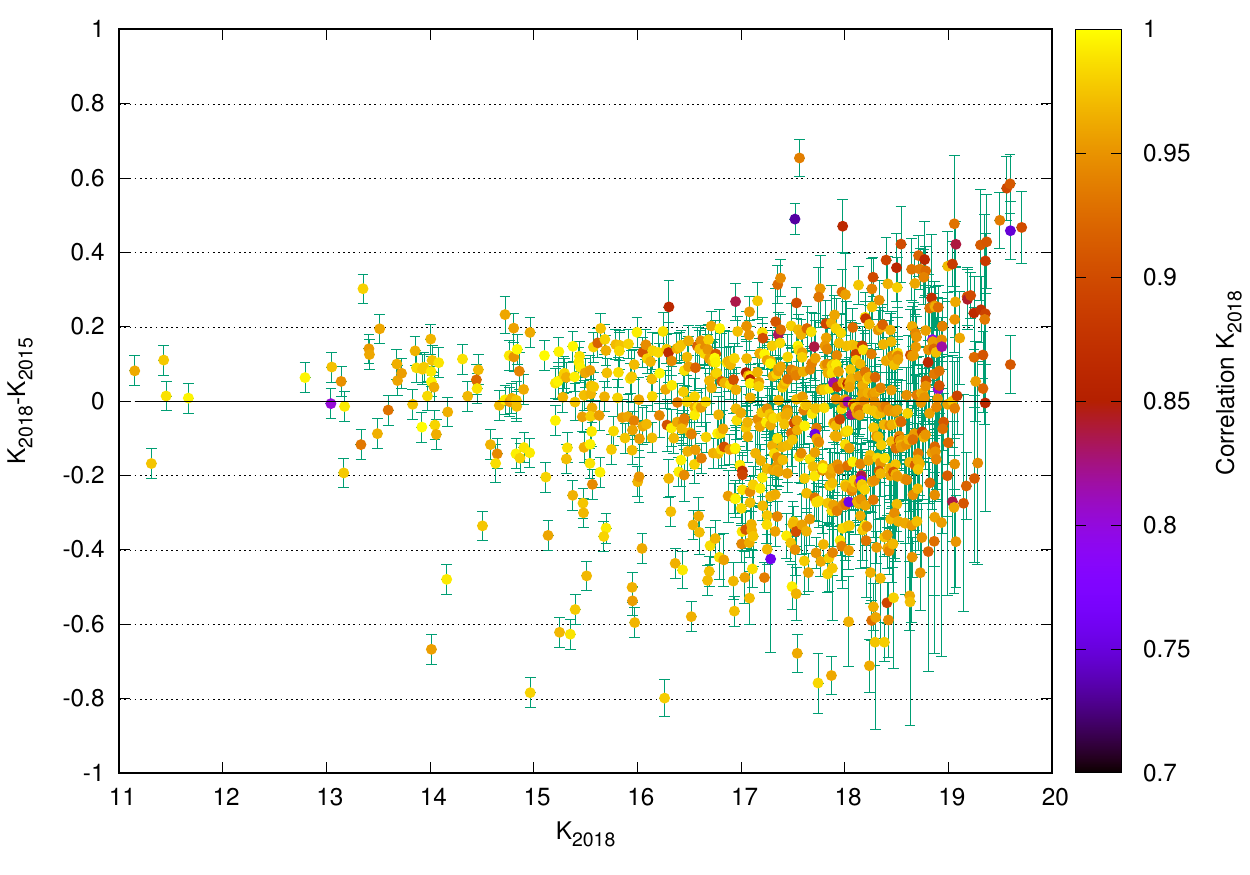}
    \caption{Difference between the K magnitude of the sources within two epochs versus their K magnitude in 2018. The colour shows the correlation between the detected source and the input PSF. The higher the value, the more similar the shape of PSF is to the reference input PSF.}
    \label{extra:KKcorr}
\end{figure}

\begin{figure}
    \centering
    \includegraphics[trim=100 0 110 0,clip,width=0.45\textwidth]{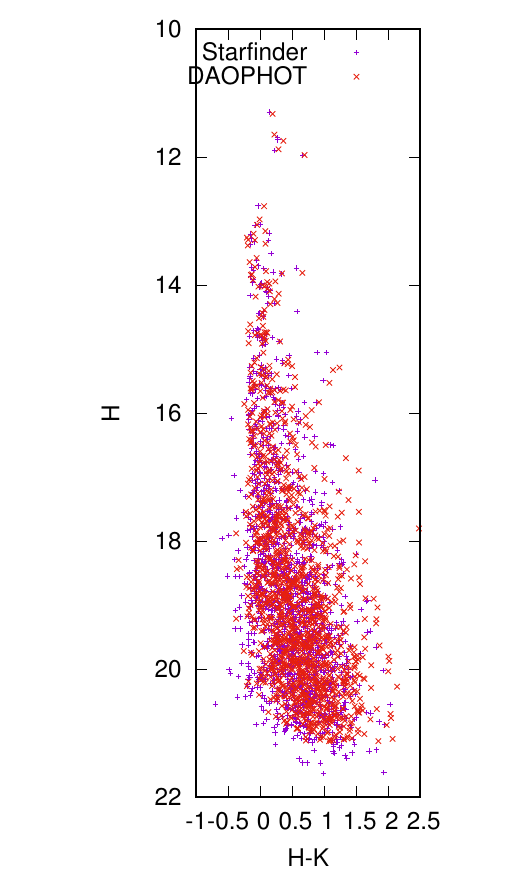}
    \includegraphics[trim=110 0 100 0,clip,width=0.45\textwidth]{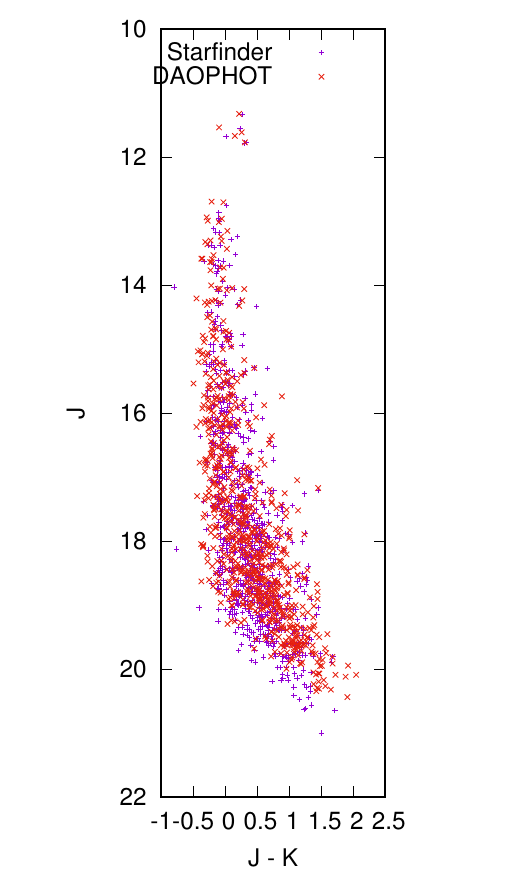}\\
    \includegraphics[]{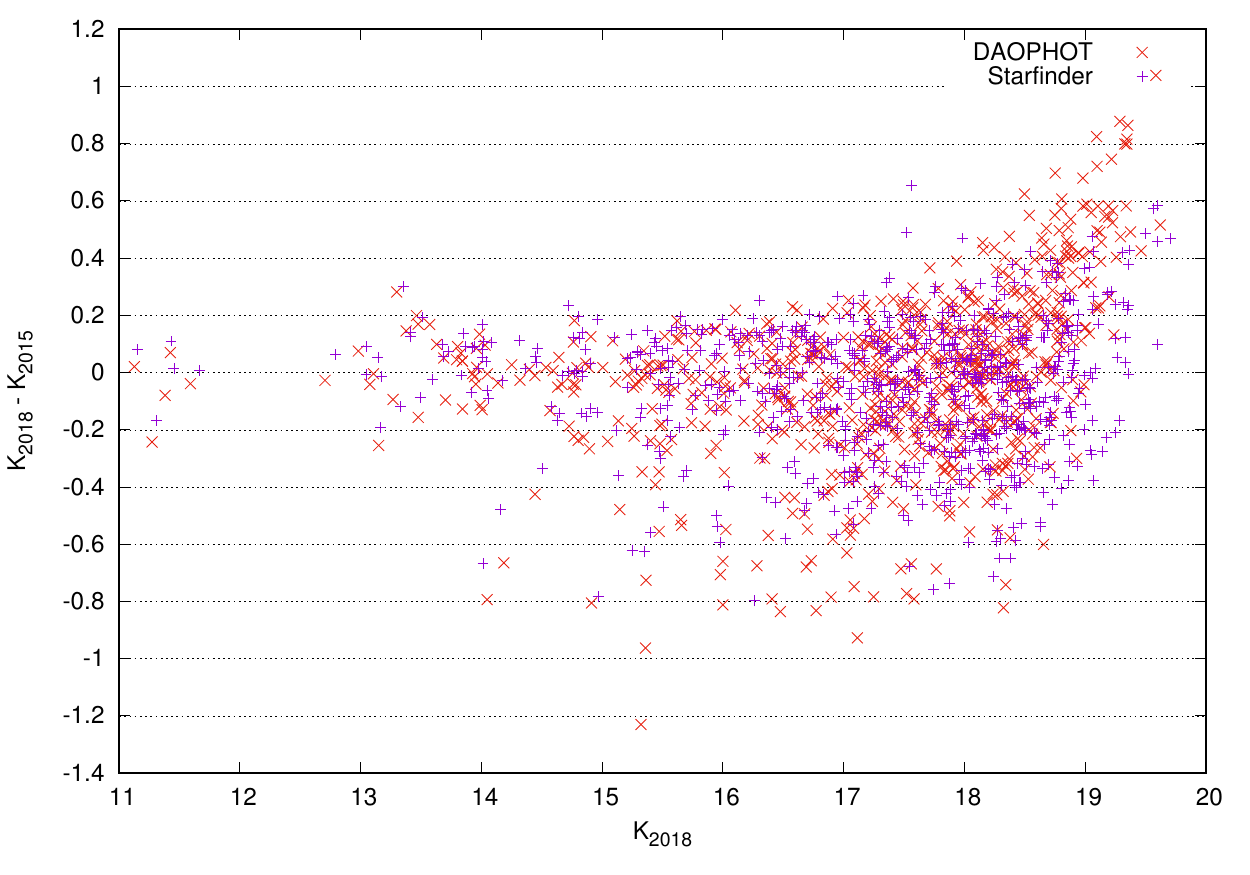}
    \caption{Comparing the photometry by {\it{Starfinder}} (purple crosses) with {\it{DAOPHOT}} (red pluses). Top show the CMD in H-K in 2018 data (left) and J-K in 2015 (right). Bottom is the comparison between the K magnitude of common sources between two epochs.}
    \label{extra:KKfind}
\end{figure}

\begin{figure}
    \centering
    \includegraphics[width=0.5\textwidth]{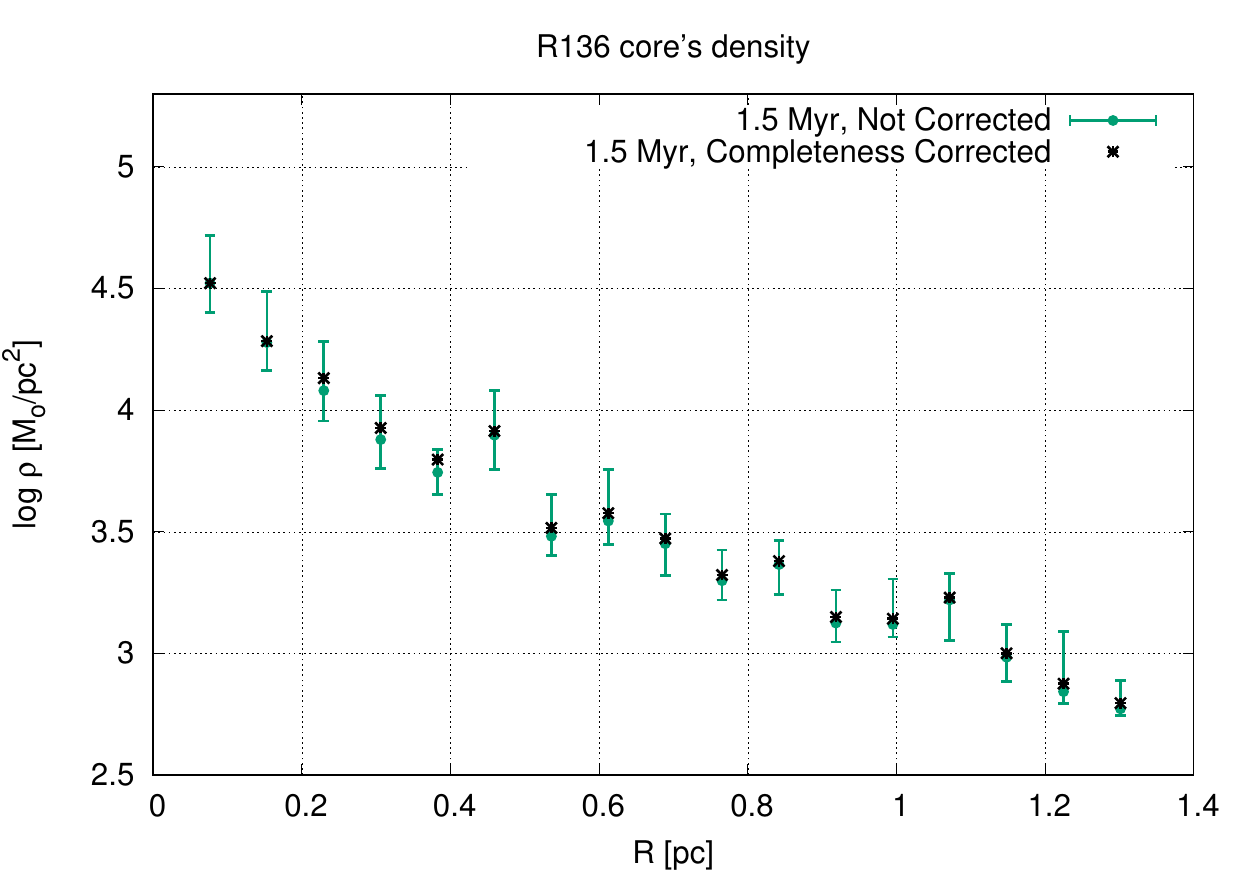}\\
    \includegraphics[width=0.5\textwidth]{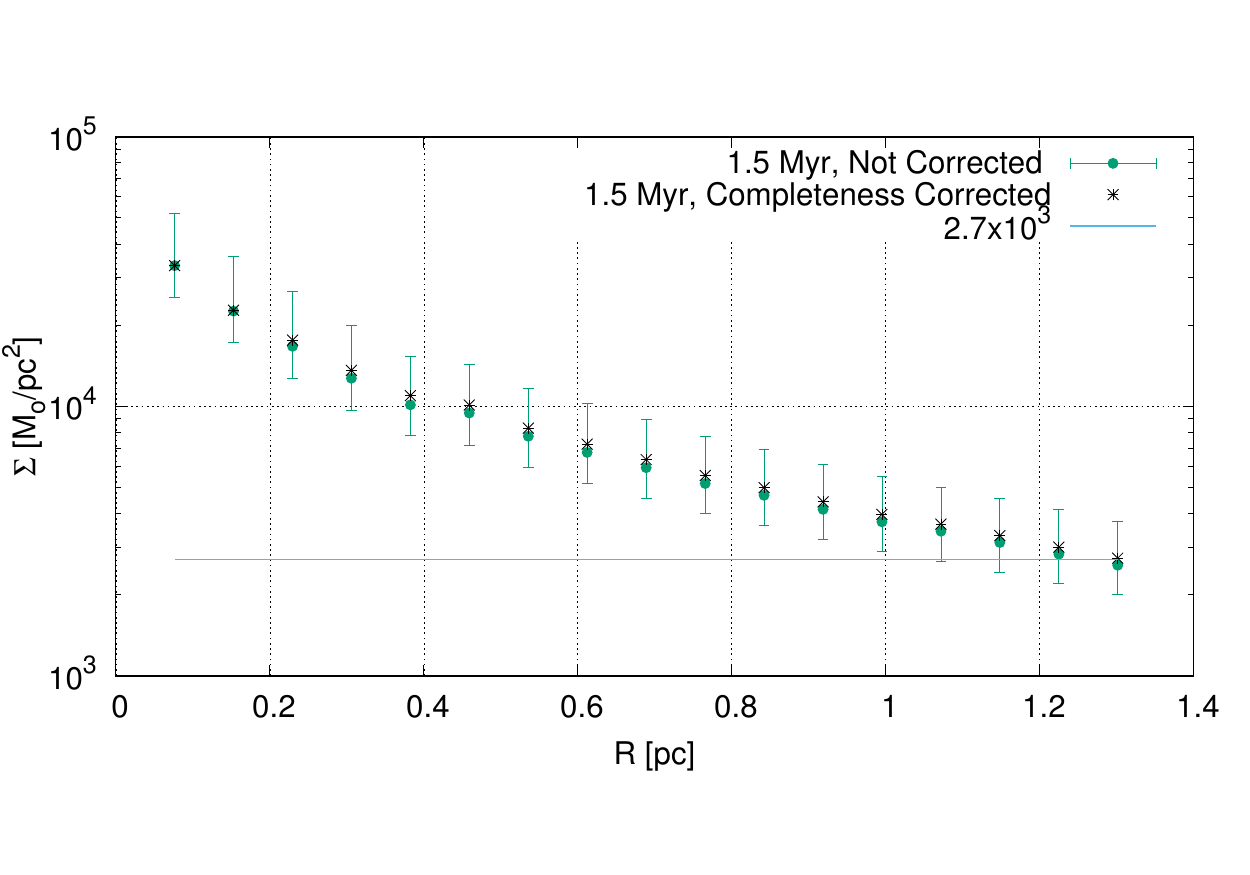}\\
    \includegraphics[width=0.5\textwidth]{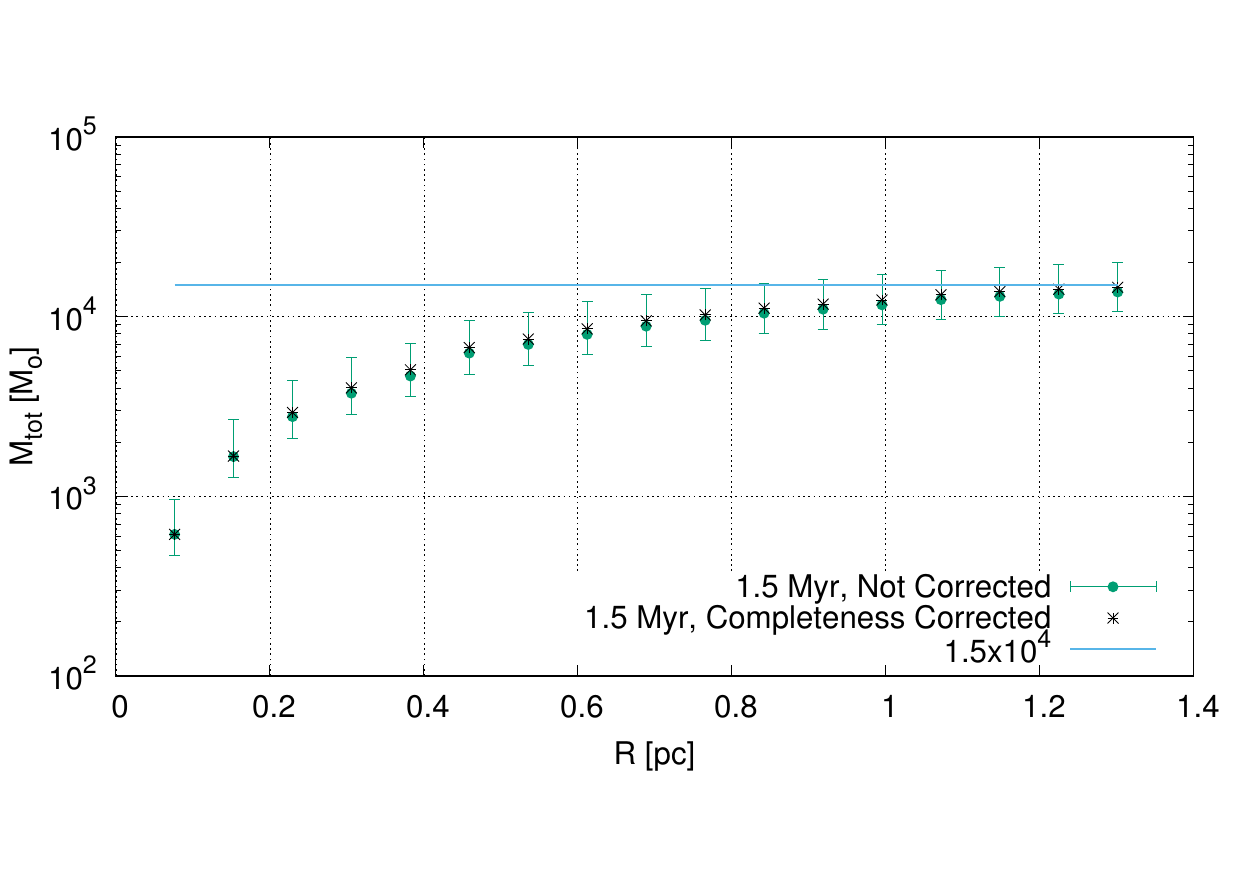}
    \caption{Comparing the effect of completeness correction on the density of the cluster at 1.5 Myr. Same as Figure \ref{fig:density} for 1.5 Myr. Black stars are the completeness corrected values.}
    \label{extra:densityCC}
\end{figure}

\begin{figure}
    \centering
    \includegraphics{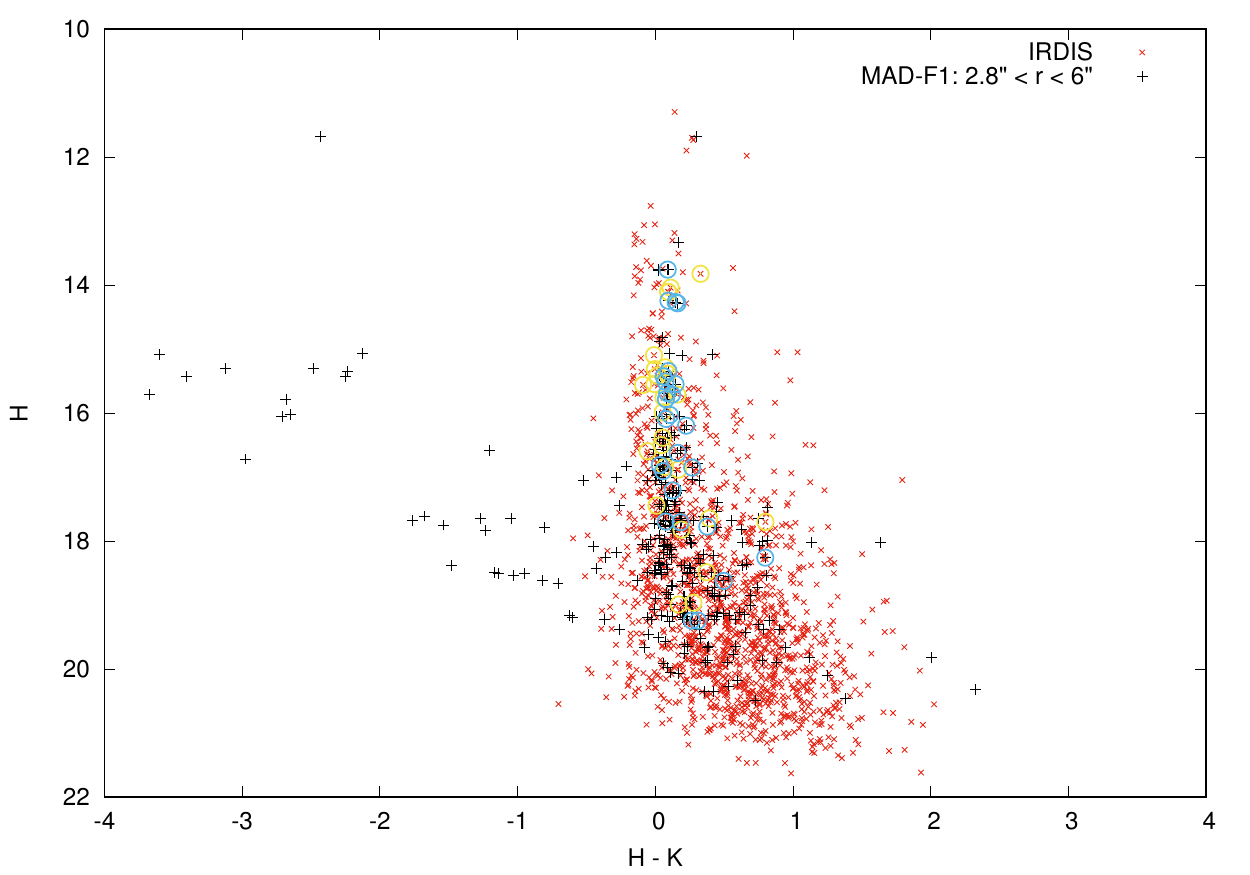}\\
    \includegraphics{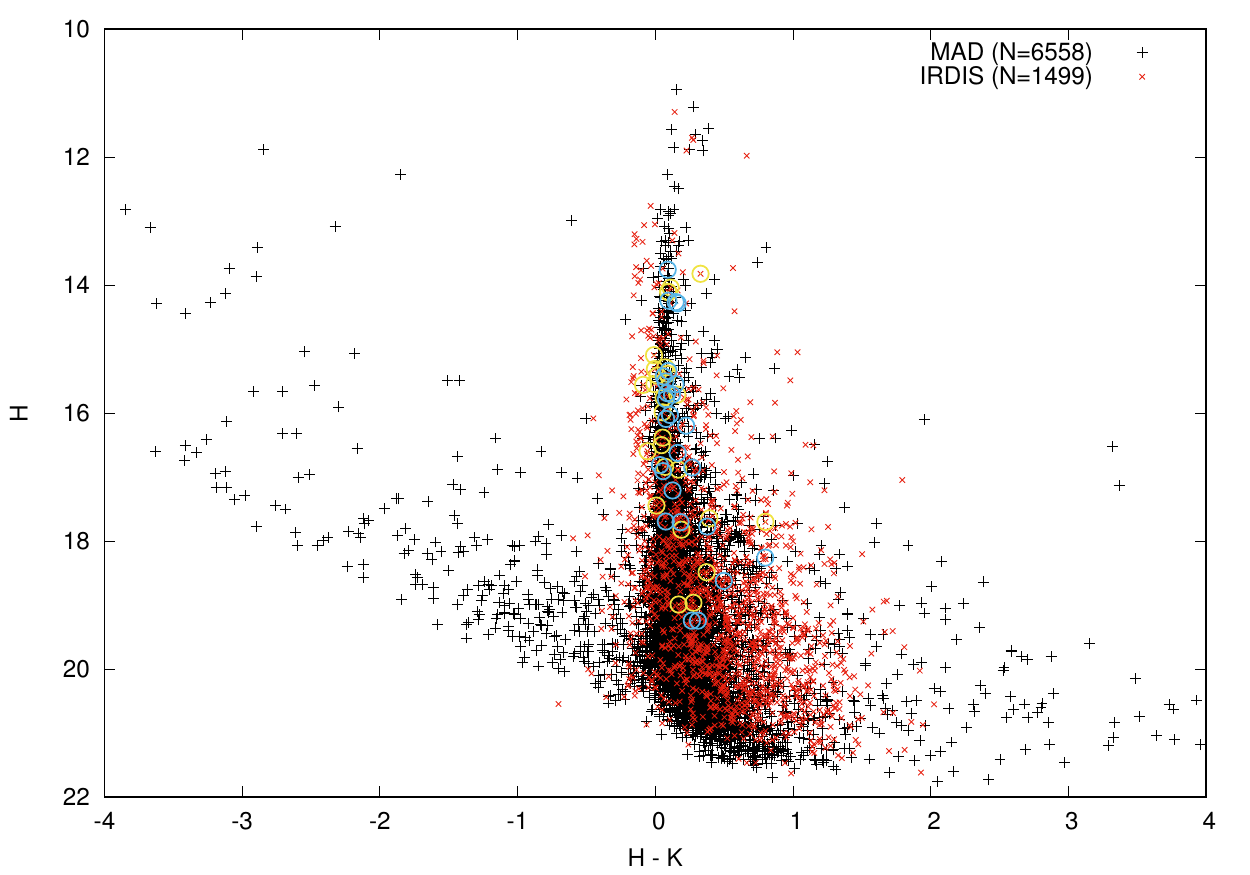}
    \caption{Comparing H-K CMD between IRDIS (this work, red crosses) and MAD (Campbell et al. (2010), black pluses). Top shows the MAD data within the same FOV of IRDIS (2.8$"$ excluded), and bottom shows the whole MAD data covering F1-F2-F3.
    yellow and blue circles, show the 28 common sources used for adjusting zeropoints.
    The number of common sources between IRDIS FOV and MAD catalogue is larger than 26, but since both our catalogues have pix position for the location of stars, we needed to identify stars by eye in each image/catalogue. 
    }
    \label{extra:mad}
\end{figure}

\onecolumn

\begin{center}
\begin{longtable}{l c c c c  c c c  c c c }\caption{Sample catalogue of detected sources in H and K data in 2018.  Full table is available online.\\
ID$_K$ is the stars identification in our K-band catalogue. K and H are the K and H magnitude of stars in 2018. Correlation$_H$ and Correlation$_K$ are the correlation between the PSF of detected source with the reference PSF in H and K images. $\sigma$ is the photometric error estimated for the positions and magnitudes.  }\\
ID$_K$&X&$\sigma_x$&Y&$\sigma_y$&K&$\sigma_K$&Correlation$_{K}$&H&$\sigma_H$&Correlation$_H$\\
&[pix]&[pix]&[pix]&[pix]&&&&&&\\
\hline
\endfirsthead
ID$_K$&X&$\sigma_x$&Y&$\sigma_y$&K&$\sigma_K$&Correlation$_{K}$&H&$\sigma_H$&Correlation$_H$\\
&[pix]&[pix]&[pix]&[pix]&&&&&&\\
\hline
\endhead
    1&  476.82&0.00&  517.37&0.00&   11.15&0.00&0.96&   11.29&0.00&0.97\\
2&  256.56&0.00&  354.97&0.00&   11.31&0.00&0.97&   11.98&0.00&0.97\\
3&  468.62&0.00&  521.19&0.00&   11.43&0.00&0.98&   11.70&0.00&0.94\\
4&  503.11&0.00&  489.78&0.00&   11.45&0.00&0.99&   11.73&0.00&0.95\\
5&  325.86&0.00&  442.90&0.00&   11.67&0.00&1.00&   11.90&0.00&0.95\\
6&  459.19&0.00&  532.97&0.00&   12.79&0.00&1.00&   12.76&0.00&0.93\\
7&  520.87&0.00&  479.45&0.00&   13.04&0.00&0.80&   13.18&0.00&0.90\\
8&  355.58&0.00&  520.89&0.00&   13.05&0.00&0.97&   13.05&0.00&0.98\\
9&  497.73&0.00&  543.72&0.00&   13.14&0.00&0.95&   13.06&0.00&0.99\\
10&  231.27&0.00&  332.67&0.00&   13.17&0.00&0.98&   13.73&0.00&0.98\\
\end{longtable}
\end{center}

\bsp	
\label{lastpage}
\end{document}